\documentclass[a4paper,11pt]{article}
\pdfoutput=1 

\usepackage{jheppub} 
\usepackage{comment}
\usepackage[T1]{fontenc} 
\usepackage{pict2e}
\usepackage{ifthen}
\usepackage[percent]{overpic}
\usepackage{array}
\usepackage{tikz}
\definecolor{oli}{RGB}{0, 132.6, 15.3}
\usepackage{amsthm}
\theoremstyle{definition}

\title{\boldmath Swampland Distance Conjecture for One-Parameter Calabi-Yau Threefolds}

\author[a]{Abhinav Joshi}
\author[a,b]{Albrecht Klemm}

\affiliation[a]{Bethe Center for Theoretical Physics\\University of Bonn, D-53113 Bonn}
\affiliation[b]{Hausdorff Center for Mathematics,\\University of Bonn, D-53113 Bonn}

\emailAdd{joshi@th.physik.uni-bonn.de}
\emailAdd{aklemm@th.physik.uni-bonn.de}

\abstract{We investigate the swampland distance conjecture (SDC) in the complex moduli space of type II compactifications on one-parameter Calabi-Yau threefolds. This class of manifolds contains hundreds of  examples and, in particular, a subset of 14 geometries with hypergeometric differential Picard-Fuchs operators. Of the four principal types of singularities that can occur --- specified by their limiting mixed Hodge structure --- only the $K$-points and the large radius points (or more generally the $M$-points) are at infinite distance and therefore of interest to the SDC. We argue that the conjecture is fulfilled at the $K$- and the $M$-points, including models with several $M$-points, using explicit calculations in hypergeometric models which contain typical  examples of  all these degenerations. Together with previous work on the large radius points,  this suggests that the SDC is indeed fulfilled for one-parameter Calabi-Yau spaces.}

\begin{document} 
\maketitle
\flushbottom

\section{Introduction and Summary}\label{sec:intro}

Quantum gravity is bound to become important in any consistent 
UV completion of physics, for which  string theory is a good 
candidate. The criteria which distinguish the vast space of semi-classically 
consistent low energy theories which admit no consistent UV completion, 
dubbed \emph{the swampland}, from the one which possibly has such a 
completion, called \emph{landscape}, have been proposed for effective  
theories, in particular, such with continuous moduli. These criteria are known as the 
\emph{swampland conjectures}~\cite{Ooguri:2006in}. While some swampland 
conjectures such as the \emph{weak gravity 
conjecture} are not specific to string theory, it has been argued 
that they are fulfilled in all known string vacua. This might reflect 
a certain bias, nevertheless it seems a reasonably general approach 
to study the border  between swampland and landscape starting 
from within the string theory moduli space. 

In this work, we will examine one of the conjectures called the 
\emph{swampland distance conjecture} (SDC)~\cite{Ooguri:2006in}. 
One starts at a point $p_0$ in the interior of the moduli space  ${\cal M}$ 
of  an effective theory, which  in the string  theory  would be defined by the 
string-- field theory  correspondence  principle.  The conjecture makes a stringent 
statement about the range of validity of this effective action. It  states that the 
theory at another point $p\in {\cal  M}$  at a distance $d(p,p_0)$, exhibits an 
infinite tower of exponentially light states whose mass scales as
\begin{equation}\label{eqn:intro_sdc}
	m_p\sim m_{p_0}e^{-\alpha d(p,p_0)}\ .
\end{equation}
Here the distance $d(p,p_0)$ is measured with the metric which fixes the kinetic terms of the moduli fields, 
$m_{p_0}$ is the mass of a given state at the point $p_0$, $m_p$ is the mass at the point $p$ and 
$\alpha\in \mathbb{R}_+$ is an a priori undetermined model dependent parameter. In the \emph{refined} swampland distance conjecture (RSDC), $\alpha$ is conjectured to be of $\mathcal{O}(1)$ in natural 
units~\cite{Klaewer:2016kiy}.  The point is that as the distance diverges, an infinite tower of 
states appears with mass below any finite mass scale used to define the effective action by integrating 
out the  UV modes. This clearly restricts the validity of the effective action  to a finite volume in ${\cal M}$.

In type IIB compactification on a family of Calabi-Yau threefolds $X_z$, the metric for an interesting class of moduli is identified with the Weil-Petersson (WP) metric of the complex structure moduli space ${\cal M}_{cs}$ parametrised by the complex moduli $z$. This K\"{a}hler metric is given exactly by the period vector $\Pi$ of the Calabi-Yau manifold in an integer symplectic basis. Also, the mass $m_\mathbf{q}$ of a BPS 3-brane state with charge vector $\mathbf{q}\in H_3(X,\mathbb{Z})$ can be calculated using the periods $\Pi$.  Moreover, mirror symmetry allows to extract exact information of the analogous data in the  complexified K\"ahler moduli space ${\cal M}_{cks}$ 
of type IIA compactifications on the mirror $Y$. Powerful mathematical tools, in particular  
the Frobenius method for constructing periods,  Schmidt's orbit theorem describing the mixed 
Hodge structure (MHS) at the singular loci $p_*$ in ${\cal M}_{cs}$  and Bridgeland condition for the stability (BS) 
of BPS states, strongly restrict the behaviour of the WP metric  and the spectrum of (light) BPS states 
at singular points of infinite distance from the interior of ${\cal M}_{cs}$.  While they are not strong 
enough to provide complete arguments to prove the SDC  on the complex moduli space 
in general, the latter makes some interesting and concrete predictions for the period 
geometry and the BPS spectrum.

It is, therefore, useful to explore SDC in a simple yet variable setting.  A very 
suitable class is that of the one parameter families of Calabi-Yau 3 folds with Betti number $b_3=4$, 
reviewed in~\cite{MR3822913}.  All known Picard-Fuchs operators ${\cal L}$, which are  of order four 
and specify, as  solutions, four periods of the holomorphic $(3,0)$ form $\Omega$, have been 
collected  in~\cite{opcylist}. The  form  of  ${\cal L}$  is very much restricted  by special geometry 
and  the possible degenerations of geometric periods at singular points $p_*$. The latter provides the distinction between points at infinite  and finite distance which is of great importance to us. This distinction is encoded  in the local monodromy $M\in {\rm Sp}(4,\mathbb{Z})$ or more precisely in the limiting 
mixed Hodge structure (LMHS) as  we will explain  in section~\ref{sec:OPMS}. There are four principle
classes of singularities: {\sl $F$-points} related to orbifold points,  {\sl $C$-points} are a 
generalisation\footnote{These generalisations can be understood as  orbifoldisations.} of conifold points,  
{\sl $K$-points} are transcendental K3 points and {\sl $M$-points} which are a  generalisation  of the large 
radius or maximal unipotent monodromy points. Out of these, as shown in section~\ref{sec:metricprop}, only the singularities at $K$- and $M$-points are of interest, as they are at infinite distance from the
interior points on the moduli space.
  
The test for the SDC on the type IIB compactifications\footnote{We use type IIB language
but by mirror symmetry generally the very same techniques translate to type IIA compactifications 
except for the orphan models in~\cite{opcylist}.} can be performed as the following. With the classification of the possible infinite distance singularities 
in~\cite{MR3822913}, it remains to understand the orbifoldisation on top of the $K$- and $M$- type. Special geometry and the form of LMHS at these singularity allows us to understand the infinite distance property, due to leading logarithmic part of period degeneration. Also, it tells us that there is a 2-dimensional sub-lattice inside the rank four lattice $\Lambda \subset  H_3(X,\mathbb{Z})$ with  an infinite number of 
primitive\footnote{At the $C$-point, the lattice is 1-dimensional and yields only one primitive 
vector. A non-primitive vector corresponds to BPS bound states at 
threshold, which are not stable. Hence there is only one massless particle, to 
be integrated out to get the Wilsonian effective action.}  charge vectors $\mathbf{q}$ that 
{\sl could} correspond to infinite number of massless BPS 3-brane states. Note that these branes wrap 3-cycles whose relative size vanishes. Their BPS masses $m_\mathbf{q}$ become indeed exponentially  
small as required  by~\eqref{eqn:intro_sdc} and with a little more effort one can see more concretely that also the RSDC holds provided these BPS states exist. 
This second  part of the argument is quite successful in establishing that 
at least the necessary conditions for the SDC hold. It  applies to the $K$- and 
$M$-points\footnote{It has been argued in~\cite{Grimm:2018cpv} that it is true 
for more general LMHS that might occur in multi-parameter Calabi-Yau families.} 
mentioned above. 

We note that even though a rank two charge lattice of light states appears at $K$- and 
$M$-points  in one moduli cases, all charges in the lattice have symplectic pairing zero so that no Argyres 
Douglas points with light magnetic and  electric charges do occur in this class. 
Such singular loci can appear if in the type IIA picture a divisor collapses, 
which is only realisable if there are at least two independent divisor classes and 
hence at least two moduli, as for  instance in the case of the $E$-string in 
elliptic fibered Calabi-Yau spaces~\cite{Klemm:1996hh}. The swampland conjectures at 
the corresponding conformal points in F-theory have been studied in~\cite{Lee:2018urn}.              

In the decisive part of the argument, one needs to establish that there is an 
infinite tower of light {\sl stable states}. In the context of the above-discussed states, one has 
to check  that  these infinite primitive charge vectors correspond to stable BPS states 
in the corresponding region of the  moduli space. Powerful techniques like the geodesic web 
condition~\cite{Klemm:1996bj} that govern the stability question for BPS states in non-gravitational 
supersymmetric theories  are not available for supergravity theories from compact 
Calabi-Yau spaces. Heuristic arguments using lines of marginal stability where a BPS 
state of charge  $\mathbf{q}$ can decay into two states with linear independent charge vectors 
$\mathbf{q}_1$ and $\mathbf{q}_2$,  e.g. for arguing  the absence of magnetic monopoles of magnetic 
charge greater than one in the asymptotic free region in~\cite{Seiberg:1994rs}, are very 
difficult to apply as the lines of marginal stability become dense.       

However, in $N=2$ supergravity in 4 dimensions, there is a Schwinger one loop diagram with 
external fields  from the gravitational sector that generates a coupling $R_-^2 F_-^{2g-2}$ 
between the anti-selfdual  parts of the  curvature $R_-$ and the anti-selfdual part of the
graviphoton field strength  $F_-$. For this diagram, at the large radius, the contributions
of BPS indices corresponding to $D2$-$D0$ brane bound states can be isolated 
and are counted by the all genus topological string partition function, which predicts  
an infinite number of non-zero indices and therefore BPS states. In particular, in the A-model, 
their vanishing is  only enforced  by the Castelnuovo conditions, leaving an infinite 
number of non-vanishing BPS states.  In the B-model, using general properties  of the 
Picard-Fuchs  equations~\cite{MR3822913} one can conclude that they lead to nontrivial 
solutions, which at least guarantee an infinite number of genus zero topological string 
invariants. Moreover, these BPS states have at the large 
radius points or more generally $M$-points the asymptotic  mass behaviour as 
in~\eqref{eqn:intro_sdc}. Mirror symmetry -- in its stronger form --- states that in 
type IIA string theory compactified to four dimensions on $Y$, the mirror family of $X$, 
one has the same BPS spectrum  from wrapping even cycles in $Y$ than IIB string compactified 
on $X$ from wrapping 3-cycles. Based on this argument, carried over to the type IIB side, with the 
corresponding identifications of BPS charges, one can argue that the SDC is valid at the large radius points in type II compactifications. Further, one can also relate it to an exact counting problem 
that has been solved to a large extent concretely in~\cite{Huang:2006hq}.

The same conclusion about the large radius point has been reached with a similar
reasoning in~\cite{Blumenhagen:2018nts}, where it is argued that an infinite tower of light Kaluza 
Klein states appear in the large radius region in type IIA theory. Maybe, one can identify these KK states with some light states on the type IIB side using mirror symmetry. Yet a different argument for an infinite tower of light states was given in~\cite{Grimm:2018,Grimm:2018cpv}.  
Here, one assumes the existence of one light stable state and creates an infinite tower of light stable 
states by a parabolic (infinite order) local monodromy around $p_*$, provided that along the 
loop in ${\cal M}$ creating the monodromy, the states building the tower  do not cross lines of marginal stability. 
It  however not clear how the argument
could explain the full and intricate spectrum of the dual $D2$-$D0$ brane states calculated in~\cite{Huang:2006hq},  
as given a state with fixed  $D2$ charge it only creates an orbit with arbitrary $D0$ brane charges. 
Moreover it only establishes that the theory is invariant under monodromy or mutation actions, 
if this action is accompanied by a relabelling of the charges $\mathbf{q}$ so that  the symplectic pairing 
$\mathbf{q}^{\text{T}}\ \Sigma\ \Pi$ is invariant. With this understanding, one does not need to create 
new states, under this action, to keep the theory invariant.

In  any case, none of the arguments stated above work at the $K$-point. The argument 
of~\cite{Blumenhagen:2018nts} fails because the $K$-points does not correspond at an 
infinite volume point in type IIA, whereas the argument of~\cite{Grimm:2018,Grimm:2018cpv} fails 
because each primitive state  with charge $\mathbf{q}$ creates only a finite orbit under 
the local monodromy. The latter is the criterion that light 
states do have to fulfil in order for a local effective action could exist according 
to~\cite{Seiberg:1994rs}.       

At the $K$-point, all  states  of the rank two lattice $\Lambda$  are not 
only in finite orbits, but their symplectic pairing is zero. The first fact
does not rule out an effective action a priori according to the argument 
of~\cite{Seiberg:1994rs} and the second fact allows one a 
Lagrangian description. One has to argue therefore that the primitive 
vectors in the lattice $\Lambda$ do exist as stable states so 
that an infinite tower of states  invalidates the effective action. Our arguments 
regarding the stability are also not complete, but highly plausible if one  follows 
the reasoning of~\cite{Banks:2010zn} which states that all  states allowed 
by the Dirac quantisation, which are locally stable, are in the spectrum.   
To establish the latter point, we analyse the lines of marginal stability for 
the exponentially light BPS states and show that, for a given charge associated to a state becoming massive at singularity, there is always a finite radius 
around the $K$- point in which an infinite number of light states are stable 
against decay into each other as well as into the above mentioned charge.    
                      
 This gives good support for the claim that the SDC holds at the $K$-points 
 due to the existence of an infinite tower of stable BPS states. 
 The fact that they are BPS states suggests that at the $K$-points 
 one has a similar interesting counting problem as at large radius that can be 
 related to higher genus string amplitudes captured by the Schwinger loop 
 amplitude, which are known at this points due to the work of~\cite{Huang:2006hq}. 
 We leave the  analysis of this problem for future work.    
 
While the $K$-point is clearly the most challenging from the point of view of 
establishing the SDC in detail, also, for $M$-point we discover a qualitatively   
new and interesting phenomenon based on our understanding of the spectrum 
of the light states. There are one parameter models which have several 
$M$-points in their moduli space. For example, four quadrics
in $\mathbb{P}^7$ called  $X_{2,2,2,2}$ is a hypergeometric case with two $M$-points in which this feature (of multiple $M$-points) 
can be studied in an easy setting. One finds 
that the different leading behaviour of the periods at the second  $M$-point 
relatively to first one can be absorbed into a K\"ahler transformation, so 
that the BPS states as captured by the topological string is the  same as at 
the first $M$-point. However, the states that become exponentially light 
at the second $M$-points and play the role of the $D2$ and $D0$ 
bound states in the Schwinger Loop  calculation do not  correspond to 
$D2$ and $D0$ branes at the first $M$-point, but rather  
to a combination involving also $D6$ and $D4$ brane charges as 
seen from the first $M$-point.

This paper is structured as follows: In section~\ref{sec:special}, we review the principle structure  of 
special geometry, which is a powerful tool for us. In section~\ref{sec:onepara}, we review first, in subsection~\ref{sec:OPMS}, the list of  one parameter Calabi-Yau manifolds and their degenerations 
leading to the four types of limiting mixed Hodge structures, up to orbifold actions. 
We  then establish the metric properties from local  considerations in subsection~\ref{sec:metricprop}.  
After a discussion of all hypergeometric cases, we introduce the geometric realisations 
of the Calabi-Yau threefolds in subsection~\ref{sec:subgeoreal}, which serve as our main examples.  Section~\ref{sec:periods} 
contains the calculations of periods on $\mathcal{M}_{cs}(X)$. We start in subsection~\ref{sec:mpoint}
to fix an integral symplectic basis near the large radius point ($\psi=\infty$). In subsection~\ref{sec:cpoint}, we calculate 
the periods near the conifold ($\psi=1$) and find the transition matrices. In subsection~\ref{sec:spoint}, we write down the periods near the third singular point ($\psi=0$), using Meijer G-functions, and the transition matrices. With the periods on the whole moduli space and the special geometry, in 
section~\ref{sec:metric}, we calculate and plot the metric on $\mathcal{M}_{cs}(X)$ over the moduli space and also give it's limiting behaviour as one approaches the singular points. 
Section~\ref{sec:sdc} contains the calculations for the SDC. We first contrast the $D2$ and $D0$ 
bound state proposal versus the Kaluza Klein proposal in subsection~\ref{sec:psi_infty}. Subsection~\ref{sec:psi_0} discusses, in detail, all the type of singularities that can occur as $\psi \rightarrow 0$ which lead to infinite distance. In particular, we focus on the lattice of exponentially light states at the $K$-point and the second\footnote{Second to the $M$-point at large radius.} $M$-point. We first work with the manifold denoted as $X_{3,3}$ in subsection~\ref{sec:X33} which has $K$-point at $\psi=0$ and after identifying the candidate stable BPS $3$-branes, in section~\ref{sec:stability1}, we study their stability properties analysing their lines of marginal stabilities. We then analyse the $X_{2,2,2,2}$ model 
in subsection~\ref{sec:X2222} which has a second $M$-point at $\psi=0$ and identify the candidate states, and in 
subsection~\ref{sec:stability2}, we analyse the lines of marginal stability. In subsection~\ref{sec:othermodels}, we 
shortly summarise the findings for the other hypergeometric one parameter models for which $K$-points
arise at $\psi=0$, for which, we have done the same analysis as for the key examples. 
Finally in Appendix~\ref{app:mgfunction}, we collect facts about the Meijer G-functions and the Barnes integral method
used to make the analysis.

\section{Special Geometry}\label{sec:special}
Let's start by reviewing some facts about the complex structure moduli space, $\mathcal{M}_{cs}(X)$ (or simply $\mathcal{M}_{cs}$), which is a special K\"{a}hler manifold~\cite{CANDELAS1991455}. Since we will be working with one-parameter spaces, we restrict to $\text{dim}(\mathcal{M}_{cs})=1$. On $\mathcal{M}_{cs}$, we can define the K\"{a}hler potential as
\begin{equation}\label{eqn:Kahler}
	K(z,\bar{z})=-\log\left(i\int_X\Omega(z)\wedge\overline{\Omega(z)}\right)\ ,
\end{equation}
where $\Omega(z)$ is the non-vanishing $(3,0)$-form on $X$ and $z$ parameterises $\mathcal{M}_{cs}$. This can be used to write down the metric on $\mathcal{M}_{cs}$,
\begin{equation}\label{eqn:Metric}
	G_{z\bar{z}}=\frac{\partial^2}{\partial z\partial\bar{z}}K(z,\bar{z})\ .
\end{equation}
The metric can be used to calculate the scalar curvature using
\begin{equation}\label{eqn:scalar}
	R=G^{z\bar{z}}R_{z\bar{z}}=G^{z\bar{z}}(-\partial_z\bar{\partial}_z\log(\text{det }G))\ ,
\end{equation}
where $G^{z\bar{z}}$ is the inverse of the metric $G_{z\bar{z}}$ and $R_{z\bar{z}}$ is the Ricci tensor.
Also, the Yukawa coupling is given by\footnote{These are given just for completeness, we will not be using these at any point in this work.}
\begin{equation}
	\kappa_{zzz}=\int_X\Omega(z)\wedge\frac{\partial^3\Omega(z)}{\partial z\partial z\partial z}\ .
\end{equation}

We now rewrite these quantities using periods. For that, we first choose an integral symplectic basis $A^I,B_I\in H_3(X,\mathbb{Z})$ in the middle homology and a dual basis $\alpha_I,\beta^I\in H^3(X,\mathbb{Z})$ in the middle cohomology with $I=0,1$,\footnote{This is due to the fact that the middle line of Hodge decomposition is: 1 1 1 1, summing up to 4, divided in a pair of 2.} such that,
\begin{equation}
\label{symplecticpairing}
    A^I\cdot B_J=-B_J\cdot A^I=\int_{X}\alpha_J\wedge\beta^I=-\int_{X}\beta^I\wedge\alpha_J=\int_{A^I}\alpha_J=\int_{B_J}\beta^I=\delta^I_J\ ,
\end{equation}
where $\delta$ is the Kronecker delta and all other pairings vanish. This allows us to define the period vector as
\begin{equation}\label{eqn:intro_period}
    \Pi=\left(
\begin{array}{c}
 F_I(z) \\
 X^I(z) \\
\end{array}
\right)=\left(
\begin{array}{c}
 \int_{B_I}\Omega(z) \\
 \int_{A^I}\Omega(z) \\
\end{array}
\right)\ .
\end{equation}
Inverting this relation, we can write
\begin{equation}
	\Omega(z)=X^I(z)\alpha_I-F_I(z)\beta^I\ .
\end{equation}
Inserting this in equation~\eqref{eqn:Kahler}, we find
\begin{equation}\label{eqn:Kahler2}
	\begin{split}
K(z,\bar{z})=&-\log\left(i\left(\overline{X^I(z)}F_I(z)-X^I(z)\overline{F_I(z)}\right)\right)\\
			=&-\log\left(-i\ \Pi(z)^{\dagger}\ \Sigma\ \Pi(z) \right)\ ,
	\end{split}
\end{equation}
where
\begin{equation}\label{eqn:sigma}
	\Sigma=\left(
	\begin{array}{cc}
	 0 & \text{Id}_{2\times 2} \\
	 -\text{Id}_{2\times 2} & 0 \\
	\end{array}
	\right)\ .
\end{equation}

Therefore, once we have the periods, we can calculate the K\"{a}hler potential using equation~\eqref{eqn:Kahler2} and then find the metric using equation~\eqref{eqn:Metric}.

\section{One-Parameter Calabi-Yau Threefolds}
\label{sec:onepara}
In this section, we will describe  the moduli geometry of complex families of Calabi-Yau threefolds  
and recall the classification and properties of their singular fibers.
\subsection{One-Parameter Families, Monodromies and  Singularities}
\label{sec:OPMS}
It is well known that the most useful classifications of these special fibers is provided 
by the monodromies of the periods of the holomorphic 3-form, which can be calculated 
from the solutions of the Picard-Fuchs operator ${\cal L}$  once an integral symplectic basis 
is provided.  The monodromies in this basis respect the symplectic pairing~\eqref{symplecticpairing}, which renders them in $Sp(4,\mathbb{Z})$. 

The coarsest classification of these types of monodromies $M$ is given by the finite order $k$ 
of the branch running into\footnote{Locally one can always choose multi-covering variables $\psi=z^\frac{1}{k}$ to remove 
the branch cut. However, globally convenient variables have the branch cuts.}  
$p_*$ and the least $n\le {\rm dim}_\mathbb{C}(X)$ determining the degree of nilpotency of $M$, i.e.
\begin{equation} 
\label{monodromy}
(M^k-{\bf{1}})^{n+1}=0 \ . 
\end{equation}   
It is convenient to apply a multiplicative  Jordan-Chevalley decomposition to $M$ and split it as
\begin{equation} 
M=U S\ ,
\end{equation}  
where $U$ and $S$ is a commuting pair of {\sl unipotent} and {\sl semi simple} matrices respectively. Special geometry 
and the fact that only regular singular points should occur restricts the form of possible differential 
operators ${\cal L}$. Moreover if one assumes a point of maximal unipotent monodromy 
(MUM), given by $k=1$ and $n=3$, then the integrality of the mirror map 
as well as of the genus zero BPS numbers yields further constraints. In particular for  
a fixed number of singularities  and apparent singularities one can provide classifications (see~\cite{opcylist} 
and~\cite{MR3822913} for a recent review). For example there are exactly $14$ hypergeometric 
cases with three singularities including a MUM point  and no apparent singularities~\cite{MR2282973}.  
Thirteen are realised geometrically by generically smooth hyper-surfaces in weighted projective spaces~\cite{MR1191426,Klemm:1992tx,Font:1992uk} 
or complete intersections in projective spaces~\cite{MR1201748} or weighted projective spaces~\cite{Klemm:1993jj}.  
A summary of these geometrical cases including the behaviour of  all genus potentials at the singularities 
can be found in~\cite{Huang:2006hq}.  

The list~\cite{opcylist} contains however  much more  general examples that have 
unsuspected  properties, like models with multiple MUM points  leading to topological  different  
mirrors  or geometric examples that have no MUM point, so called orphans,  
leading to families with no geometric mirror. 

There are finer types of invariants than the $k$ and $n$ in~\eqref{monodromy} associated to the singular points $p_*$, in particular using  the 
nilpotent operator
\begin{equation} 
N=\log(U)\ ,
\end{equation} 
one can define in addition to the $F^p=\oplus_{i\geq p} H^{i,3-i}(X)$  filtration, a second $W_*$ filtration 
for the limiting Hodge structure that refines nilpotency index $n$ into the information in the limiting 
mixed Hodge structure (LMHS). In the following, we mention some basic definitions\footnote{See~\cite{MR3822913} for an explanation in the one-parameter 
case and~\cite{MR1677117} for the general theory for physicist and  further references.} that leads 
to the Hodge diamonds shown below.  
The  $W_*$ filtration $W_0\subset W_1\subset \ldots \subset W_{2n-1}\subset W_{2n}=H^3_{\rm lim}(X_s)$  
at the singular point $s$ is given by the action of $N$ as 
$W_0={\rm Im}(N^n)$,  $W_1={\rm Im}(N^{n-1}) \cap {\rm ker}(N)$, $W_2={\rm Im}(N^{n-2}) \cap {\rm ker}(N)+ {\rm Im}(N^{n-1}) \cap {\rm ker}(N^2)$, 
$\cdots$, $W_{2n-1}=  {\rm ker}(N^n)$. Note that $N(W_k)\subset W_{k-2}$. On the quotient spaces
${\rm Gr}_k=W_{k}/W_{k-2}$, the $N^k$ induce  an isomorphism $N^k:{\rm Gr}_{n+k}  \rightarrow {\rm Gr}_ {n-k}$ and $F^*_{\rm lim}$  
defines a rank $k$ Hodge structure on the spaces ${\rm Gr}_{k}$. The Hodge diamonds of the LMHS shows 
these ${\rm Gr}_{k}$, with $k=0,\ldots, 2n-1$ from bottom to top and for each  ${\rm Gr}_{k}$, the Hodge 
decomposition. From above it follows that $N$ acts by a shift of two rows and the original 
entries can only be shifted up or down along SW-NE lines.

For one-parameter CY families, the  LMHS  is, to a large extent, already captured by the local 
exponents  $(a_1, a_2, a_3, a_4)$ with $a_i\in \mathbb{Q}$ at the singular point $p_*$ at $x_*=0$. These exponents  
are determined  by solving for $a$ with multiplicities in ${\cal L}(x_*) x_*^a \sum_{k=0}^\infty c_k x_*^k=0$, 
where $x_*$ is a  local complex structure coordinate. They are summarised in the columns of the Riemann 
Symbol associated to ${\cal L}$ , see e.g.~\eqref{eqn:RSym} for the hypergeometric cases. For this reason, it is convenient to 
index  the ${\cal L}$ on~\cite{opcylist} by their local exponents.   The most well known  singularities, 
the MUM points and the conifold points with local exponents  $(0,0,0,0)$ and $(0,1,1,2)$ 
respectively~\eqref{eqn:RSym} occur in thousands  of different ${\cal L}$ often with 
higher multiplicity then one.  

However, in total there are only $123$ different local exponents in~\cite{opcylist}. 
For $57$ of them, the difference between different $a_i$ is either zero or a non-vanishing 
rational number (not in $\mathbb{Z}$). In this case one can infer the local types 
immediately from the local exponents and these singularities fall into four principal types. 
\begin{itemize}
\item  $F$:  $M$ is of finite order $k$ if the exponents are all different $(a,b,c,d)$. The LMHS diamond is 
$$
\begin{array}{ccccccc}
&\phantom{000}&\phantom{000}&\phantom{0}0 \phantom{0}& \phantom{000}&\phantom{000}&\\
&&0&&0&&\\
&0&&0&&0&\\
1&&1&&1&&1\\
&0&&0&&0&\\
&&0&&0&&\\
&&&0&&&\\
\end{array}
$$
\item  $C$:  $M$ has  one $2\times 2$ Jordan block  of infinite order if the exponent structure is  $(a,b,b,c)$, and the LMHS diamond is
$$
\begin{array}{ccccccc}
&\phantom{000}&\phantom{000}&\phantom{0}0 \phantom{0}& \phantom{000}&\phantom{000}&\\
&&0&&0&&\\
&0&&1&&0&\\
1&&0&&0&&1\\
&0&&1&&0&\\
&&0&&0&&\\
&&&0&&&\\
\end{array}
$$
\item  $K$:  $M$ has two such Jordan blocks if the  exponent structure  is $(a,a,b,b)$ and 
the LMHS diamond is 
$$
\begin{array}{ccccccc}
&\phantom{000}&\phantom{000}&\phantom{0}0 \phantom{0}& \phantom{000}&\phantom{000}&\\
&&0&&0&&\\
&1&&0&&1&\\
0&&0&&0&&0\\
&1&&0&&1&\\
&&0&&0&&\\
&&&0&&&\\
\end{array}
$$
\item  $M$:  $M$  has a $4\times 4$ Jordan block of infinite order if the exponents are all equal, i.e. $(a,a,a,a)$ and 
the LMHS diamond is 
$$
\begin{array}{ccccccc}
&\phantom{000}&\phantom{000}&\phantom{0}1\phantom{0}& \phantom{000}&\phantom{000}&\\
&&0&&0&&\\
&0&&1&&0&\\
0&&0&&0&&0\\
&0&&1&&0&\\
&&0&&0&&\\
&&&1&&&\\
\end{array}
$$
\end{itemize} 
We note that the singular points of type  $F$ and $C$ are at finite distance  as measured with the WP 
metric, while  type $K$ and $M$ are at infinite distance.  This follows simply from the 
logarithmic structure of the solutions associated with the Jordan block of infinite order as we will see in the next subsection.  
In all cases, $k$ in equation~\eqref{monodromy} is the LCM of the denominator of the $a_i$. The 66 remaining exponents 
have integer differences between the different $a_i$. They exhibit the same structure of the local 
exponents  above and the same Jordan blocks {\sl can} occur but the question {\sl if} 
they actually occur requires  next to leading order  study of the actual solutions to ${\cal L}$. For example, the conifold is indeed 
of type $C$.  Note  that the pattern $(a,b,b,b)$ with  a $3\times 3$ Jordan block of finite or infinite order can be excluded by 
special geometry.
\subsection{Metric Properties at the Critical Points}
\label{sec:metricprop}
In the following, we would like to comment on the distance to the different types of singular points that can occur. We give an overview  
of the general structure, while explicit calculation for concrete models  can be found in later sections. For the 14 hypergeometric families
with $\psi$ parametrising the moduli space. Table~\ref{tab:allmetric2} shows the type of the singular point at $\psi=0$\footnote{See section~\ref{sec:subgeoreal} for notation of manifold and relation of $\psi$ and $z$, where $z$ appears in the Picard-Fuchs equation~\eqref{eqn:sth_pfeqn}.}, the name of the 
threefold and the exponents near $\psi=0$ in the first three columns. The fourth column has $\mu$ which is used in the next sections and it appears in the 
Picard-Fuchs equation.  The fifth and sixth column contains topological 
numbers\footnote{The calculation for these topological numbers can be found in~\cite{Hosono:1994ax}.} 
$\kappa=\int_X\omega^3=D^3$ and $c_2\cdot D=\int_X c_2(TX)\wedge\omega$ required for finding periods 
in symplectic basis and most importantly, column seven has the metric structure as one approaches the singular 
point (at $\psi=0$) and the eighth column has the distance measured from a nonsingular point to $\psi=0$. 
\begin{table}[]
\centering
\begin{tabular}{|c|c|m{2.4cm}|c|c|c|m{3.3cm}|c|}
\hline
Type & Threefold & $(a_1,a_2,a_3,a_4)$ & $\mu$ & $\kappa$ & \footnotesize$c_2\cdot D$\normalsize & Metric & Dist.\\ \hline
$M$ & $X_{2,2,2,2}(1^8)$ & $\left(\frac{1}{2},\frac{1}{2},\frac{1}{2},\frac{1}{2}\right)$ & $2^8$ & $16$ & $64$ & \vspace{1mm}$\frac{3}{4|\psi|^2\log^2(|\psi|)}$\vspace{1mm} & $\infty$  \\ \hline
$F$ &$X_{4,3}(1^52)$ & $\left(\frac{1}{4},\frac{1}{3},\frac{2}{3},\frac{3}{4}\right)$ &  $2^6 3^3$ & $6$ & $48$ & \vspace{1mm}$\frac{3^3 \Gamma \left(\frac{1}{3}\right)^6 \Gamma
   \left(\frac{3}{4}\right)^4}{2 \pi  \Gamma \left(\frac{1}{4}\right)^8{|\psi|}}$\vspace{1mm} & $<\infty$ \\ \hline
$C$ &$X_{4,2}(1^6)$ & $\left(\frac{1}{4},\frac{1}{2},\frac{1}{2},\frac{3}{4}\right)$ & $2^{10}$ & $8$ & $56$ & \vspace{1mm}$-\frac{2 3^3 \Gamma \left(\frac{3}{4}\right)^{12}|\psi|\log(|\psi|)}{\pi^6}$\vspace{1mm} & $<\infty$ \\ \hline
$F$ &$X_{5}(1^5)$ & $\left(\frac{1}{5},\frac{2}{5},\frac{3}{5},\frac{4}{5}\right)$ & $5^5$ & $5$ & $50$ & \vspace{1mm}$\frac{5^2 \Gamma \left(\frac{2}{5}\right)^5 \Gamma \left(\frac{4}{5}\right)^5}{\Gamma \left(\frac{1}{5}\right)^5 \Gamma \left(\frac{3}{5}\right)^5}$\vspace{1mm} & $<\infty$ \\ \hline
$K$ &$X_{3,3}(1^6)$ & $\left(\frac{1}{3},\frac{1}{3},\frac{2}{3},\frac{2}{3}\right)$ & $3^6$ & $9$ & $54$ & \vspace{1mm}$\frac{1}{4|\psi|^2\log^2(|\psi|)}$\vspace{1mm} & $\infty$ \\ \hline
$K$ &$X_{4,4}(1^42^2)$ & $\left(\frac{1}{4},\frac{1}{4},\frac{3}{4},\frac{3}{4}\right)$ & $2^{12}$ & $4$ & $40$ & \vspace{1mm}$\frac{1}{4|\psi|^2\log^2(|\psi|)}$\vspace{1mm} & $\infty$ \\ \hline
$C$ &$X_{3,2,2}(1^7)$ & $\left(\frac{1}{3},\frac{1}{2},\frac{1}{2},\frac{2}{3}\right)$ & $2^43^3$ & $12$ & $60$ & \vspace{1mm}$-\frac{7^3\Gamma \left(\frac{5}{6}\right)^9|\psi|^{1/3}\log(|\psi|)}{2^{8/3}\pi^{9/2}}$\vspace{1mm} & $<\infty$ \\ \hline
$C$ &$X_{6,2}(1^53)$ & $\left(\frac{1}{6},\frac{1}{2},\frac{1}{2},\frac{5}{6}\right)$ & $2^83^3$ & $4$ & $52$ & \vspace{1mm}$-\frac{3^3 \Gamma \left(\frac{5}{6}\right)^9|\psi|^{2}\log(|\psi|)}{2^{1/3}\pi^{9/2}}$\vspace{1mm} & $<\infty$ \\ \hline
$F$ &$X_{6}(1^42)$ & $\left(\frac{1}{6},\frac{1}{3},\frac{2}{3},\frac{5}{6}\right)$ & $2^43^6$ & $3$ & $42$ & \vspace{1mm}$\frac{100 \sqrt[3]{2} \pi ^{7/2}}{\sqrt{3}  \Gamma \left(\frac{1}{6}\right)^5 \Gamma \left(\frac{1}{3}\right)^2|\psi|^{1/3}}$\vspace{1mm} & $<\infty$ \\ \hline
$F$ &$X_{8}(1^44)$ & $\left(\frac{1}{8},\frac{3}{8},\frac{5}{8},\frac{7}{8}\right)$ & $2^{16}$ & $2$ & $44$ & \vspace{1mm}$0.0815626|\psi|^{1/2}$\vspace{1mm} & $<\infty$ \\ \hline
$F$ &$X_{6,4}(1^32^23)$ & $\left(\frac{1}{6},\frac{1}{4},\frac{3}{4},\frac{5}{6}\right)$ & $2^{10}3^3$ & $12$ & $32$ & \vspace{1mm}$\frac{0.0103506}{|\psi|}$\vspace{1mm} & $<\infty$\\ \hline
$F$ &$X_{10}(1^32,5)$ & \vspace{1mm}$\left(\frac{1}{10},\frac{3}{10},\frac{7}{10},\frac{9}{10}\right)$\vspace{1mm} & $2^85^5$ & $1$ & $34$ & $0.0424127$ & $<\infty$ \\ \hline
$K$ &$X_{6,6}(1^22^23^2)$ & $\left(\frac{1}{6},\frac{1}{6},\frac{5}{6},\frac{5}{6}\right)$ & $2^83^6$ & $1$ & $22$ & \vspace{1mm}$\frac{1}{4|\psi|^2\log^2(|\psi|)}$\vspace{1mm} & $\infty$ \\ \hline
$F$ &$X_{2,12}(1^44,6)$ & $\left(\frac{1}{12},\frac{5}{12},\frac{7}{12},\frac{11}{12}\right)$ & $2^{12}3^6$ & $1$ & $46$ & \vspace{1mm}$\frac{9 \Gamma \left(\frac{5}{6}\right)^6|\psi|^2}{8 \sqrt[3]{2} \pi ^3}$\vspace{1mm} & $<\infty$ \\ \hline
\end{tabular}
\caption{Showing the metric as $\psi\rightarrow 0$ on moduli space of 14 hypergeometric one parameter Calabi-Yau manifolds. Also, the distance to $\psi=0$.}
\label{tab:allmetric2}
\end{table}

First, let us show that  the  $M$- and $K$-points are at infinite distance and all 
others are at finite distance. For this we will be using the Schmidt's orbit theorem~\cite{MR0382272} which allows us to separate the part responsible for monodromy transformation and a monodromy invariant power series. To simplify the analysis, we perform coordinate transformation $\tilde{x}=x$ where $\tilde{x}$ is the local coordinate and $k$ is the LCM of denominator of $a_i$. This transformation gets rid of the semi simple part of the monodromy and allows us to write the period as
\begin{equation}\label{orbitthm}
	\Pi=\exp\left(\frac{1}{2 \pi i}\log(x)N\right)\mathbf{A}(x)\ ,
\end{equation}
where $\mathbf{A}(x)$ is a holomorphic function which can be expanded as
\begin{equation}
	\mathbf{A}(x)=\mathbf{a}_0+\mathbf{a}_1x+\mathbf{a}_2x^2+\ldots
\end{equation}
and $N=\log(U)$ where $U$ is the unipotent part of the monodromy matrix obtained by the transformation $\Pi(x)\rightarrow\Pi(e^{2\pi i}x)=U\Pi(x)$. Since $U$ is unipotent, $N$ is nilpotent meaning that $N^{i-1}\neq 0$ and $N^i=0$ for some $i\in\mathbb{Z}^{+}$. For $F$-points, $i=1$, for $C$- and $K$-points, $i=2$ and for $M$-points, $i=4$. Before writing the K\"{a}hler potential, we define another coordinate $t=\frac{1}{2\pi i}\log(x)=v+iy$ (not the mirror map).

The K\"{a}hler potential can then be written as
\begin{equation}
	\begin{split}
		K\left(t,\bar{t}\right)&=-\log\left(-i \mathbf{A}(t)^{\dagger}e^{\bar{t}N^\text{T}}\Sigma e^{tN}\mathbf{A}(t) \right)\\
						&=-\log\left(-i\mathbf{A}(t)^{\dagger}\Sigma e^{(t-\bar{t})N} \mathbf{A}(t)\right)\\
						&=-\log\left(-i (\mathbf{a}_0^\dagger+\mathbf{a}_1^\dagger e^{-2\pi i \bar{t}}+\ldots)\Sigma e^{(t-\bar{t})N}(\mathbf{a}_0+\mathbf{a}_1 e^{2\pi i t}+\ldots)\right)\ ,
	\end{split}
	\end{equation}
	where in second line, we have used the relation $N^{\text{T}}\Sigma=-\Sigma N$ which can be deduced using $U^\text{T}\Sigma U=\Sigma$. For the periods near $x=0$ or $t=i\infty$, one finds that $\mathbf{a}_i=0$ for $i<m\equiv\text{min}\{ka_1,ka_2,ka_3,ka_4\}$, which can be understood from the local solutions around the singular point. Hence, we find that
\begin{equation}
	\begin{split}
		K\left(t,\bar{t}\right)&=-\log\left(-i\mathbf{a}_m^\dagger\Sigma e^{2i yN}\mathbf{a}_m e^{-4\pi m y}+\mathcal{O}\left(e^{-4\pi my-2\pi y}\right)\right)\\
								&=4\pi m y-\log\left(-i\mathbf{a}_m^\dagger\Sigma e^{2i yN}\mathbf{a}_m +\mathcal{O}\left(e^{-2\pi y}\right)\right)\ .
	\end{split}
	\end{equation}
The first term above can be removed using a K\"{a}hler transformation $K(t,\bar{t})\rightarrow K(t,\bar{t})-f(t)-\bar{f}(\bar{t})$ with $f(t)=-2\pi mit$ and $\bar{f}(\bar{t})=2\pi mi\bar{t}$. Therefore, we get
\begin{equation}
	\begin{split}
		K\left(t,\bar{t}\right)&=-\log\left(-i\mathbf{a}_m^\dagger\Sigma e^{2i yN}\mathbf{a}_m +\mathcal{O}\left(e^{-2\pi y}\right)\right)\\
								&=-\log\left(p( y)+\mathbf{h}\right)\ , 
	\end{split}
\end{equation}
where $p( y)$ is a polynomial of degree $d$ since $N$ is nilpotent. 
Note that for $ y\rightarrow\infty$, the $\mathbf{h}=\mathcal{O}(e^{-2\pi y})$ term 
and all of it's partial derivatives are exponentially suppressed. Using this, one can readily calculate the metric as
\begin{equation}
	\begin{split}
	G=\partial_t\bar{\partial}_{\bar{t}}K(t,\bar{t})&=\frac{1}{4}\left(\partial_{ y}\partial_{ y}+\partial_{v}\partial_{v}\right)K(v,y)\\
	&=-\frac{1}{4}\frac{\left(p( y)''+\mathbf{h}\right)\left(p( y)+\mathbf{h}\right)-(p( y)'+\mathbf{h})^2}{(p( y)+\mathbf{h})^2}\\
	&=-\frac{1}{4}\frac{p(y)''p(y)-p(y)'^2+\mathbf{h}}{p^2+\mathbf{h}}\sim\frac{1}{4}\frac{p'^2-p''p}{p^2}+\mathbf{h}\\
	&\sim\frac{1}{4}\frac{d^2-d(d-1)}{y^2}+\mathbf{h}=\frac{d}{y^2}+\mathbf{h}\ .
	\end{split}
\end{equation}
Therefore, for $d=0$, we get finite distance on integrating the metric distance from some non singular point to $y=\infty$, whereas, 
for $d>0$, the distance diverges logarithmically as $y\rightarrow\infty$. 
As noticed in~\cite{wang1997incompleteness}, one can conclude that the 
degree of the polynomial $d=\text{deg}\ p( y)=\text{max}\{l,N^l\mathbf{a}_m\neq0\}$.

In the following, we will be using Frobenius basis instead of symplectic basis. Let's denote the Frobenius basis by $\tilde{\Pi}$ and the transformation matrix by $T$ such that
\begin{equation}
	\Pi=T\tilde{\Pi}=T\exp\left(\frac{1}{2\pi i}\log(x)\tilde{N}\right)\tilde{\mathbf{A}}(x)\ ,
\end{equation}
which, using equation~\eqref{orbitthm}, can be rewritten as
\begin{equation}
	\begin{split}
&\exp\left(\frac{1}{2\pi i}\log(x)N\right)\mathbf{A}(x)=T\exp\left(\frac{1}{2\pi i}\log(x)\tilde{N}\right)\tilde{\mathbf{A}}(x)\\
			\implies &\exp\left(\frac{1}{2\pi i}\log(x)N\right)\mathbf{A}(x)=\exp\left(\frac{1}{2\pi i}\log(x)T\tilde{N}T^{-1}\right)T\tilde{\mathbf{A}}(x)\ .
	\end{split}
\end{equation}
On comparing the two sides, we can read off that
\begin{equation}
	N=T\tilde{N}T^{-1},\quad \mathbf{A}(x)=T\tilde{\mathbf{A}}(x)
\end{equation}
and hence, the statement for the degree $d$ translates to $d=\text{deg}\ p(y)=\text{max}\{l,\tilde{N}^l\tilde{\mathbf{a}}_m\neq 0\}$.

Let's look at each type of singular point:
\begin{itemize}
	\item $F$-point $(a,b,c,d)$: For this type, we know that $\tilde{N}=0$ ($N=0$) from above, hence $d=0$, making it a finite distance point.
	\item $C$-point $(a,b,b,c)$: For this type, with $a<b<c$, the structure of exponents locally allow us to write the periods in Frobenius basis as
	\begin{equation}
		\tilde{\Pi}=\left(\begin{array}{c}
			\nu_0(x)\\
			\nu_0(x)\log(x)+s_0(x)\\
			s_1(x)\\
			s_2(x)\\
		\end{array}\right),\text{ with }\tilde{\mathbf{a}}_m=\left(\begin{array}{c}
			0\\0\\\alpha \\0
		\end{array}\right)\ .
	\end{equation}
Here $\nu_0(x)$ and $s_0(x)$ are power series with leading term of  order $x^{kb}$, $s_1(x)$ and $s_2(x)$ are power series with leading term of order $x^{ka}$ and $x^{kc}$ 
respectively and $\alpha\in\mathbb{C}$. Note that here $m=ka$. In this basis, the unipotent part of the monodromy matrix and it's logarithm can be written as
	\begin{equation}
		\tilde{U}=\left(\begin{array}{cccc}
			1 & 0 & 0 & 0\\
			2\pi i & 1 & 0 & 0\\
			0 & 0 & 1 & 0\\
			0 & 0 & 0 & 1\\
		\end{array}\right),\quad\tilde{N}=\log(\tilde{U})=\left(
\begin{array}{cccc}
 0 & 0 & 0 & 0 \\
 2 i \pi  & 0 & 0 & 0 \\
 0 & 0 & 0 & 0 \\
 0 & 0 & 0 & 0 \\
\end{array}
\right)\ .
	\end{equation}
Note that this monodromy is due to the logarithmic term in the period vector $\tilde{\Pi}$. The contribution to the
monodromy from power series due to the branch cuts in the power series shows up in semi-simple part and is irrelevant for 
the argument about the metric. Clearly, for the above  $\tilde{N}\tilde{\mathbf{a}}_m=0$. Hence, $d=0$, making the $C$ point a finite distance 
point.
	\item $K$-point $(a,a,b,b)$: For this point, with $a<b$, the exponents imply that the solutions are given by
\begin{equation}
	\tilde{\Pi}=\left(\begin{array}{c}
		\nu_0(x)\\
		\nu_0(x)\log(x)+s_0(x)\\
		\nu_1(x)\\
		\nu_1(x)\log(x)+s_1(x)\\
	\end{array}\right),\text{ with }\tilde{\mathbf{a}}_m=\left(\begin{array}{c}
		\alpha\\\alpha'\\0\\0
	\end{array}\right)\ ,
\end{equation}
where $\nu_0(x)$ and $s_0(x)$ are power series with leading term of order $x^{ak}$, $\nu_1(x)$ and $s_1(x)$ have leading term order $x^{bk}$ and $\alpha,\alpha'\in\mathbb{C}$. Here, $m=ak$. Again, the unipotent part of the monodromy and it's logarithm will be
	\begin{equation}
		\tilde{U}=\left(\begin{array}{cccc}
			1 & 0 & 0 & 0\\
			2\pi i & 1 & 0 & 0\\
			0 & 0 & 1 & 0\\
			0 & 0 & 2\pi i & 1\\
		\end{array}\right),\quad\tilde{N}=\log(\tilde{U})=\left(
\begin{array}{cccc}
 0 & 0 & 0 & 0 \\
 2 i \pi  & 0 & 0 & 0 \\
 0 & 0 & 0 & 0 \\
 0 & 0 & 2 i \pi  & 0 \\
\end{array}
\right)\ .
	\end{equation}
Note that now 
\begin{equation}
	\tilde{N}\tilde{\mathbf{a}}_m=\left(\begin{array}{c}
		0\\2\pi i \alpha \\0\\0
	\end{array}\right), \quad\text{ and }\quad
	\tilde{N}^2\tilde{\mathbf{a}}_m=0\ .
\end{equation}
Hence, we can conclude that $d=1$ for $K$ points, making it an infinite distance point.
\item $M$-point $(a,a,a,a)$: Here, exponents lead us to the following period vector
\begin{equation}
	\tilde{\Pi}=\left(\begin{array}{c}
		\nu_0(x)\\
		\nu_0(x)\log(x)+s_0(x)\\
		\nu_0(x)\log^2(x)+2s_0(x)\log(x)+s_1(x)\\
		\nu_0(x)\log^3(x)+3s_0(x)\log^2(x)+3s_1(x)\log(x)+s_2(x)\\
	\end{array}\right),\text{ with }\tilde{\mathbf{a}}_m=\left(\begin{array}{c}
		\alpha\\\alpha'\\\alpha''\\\alpha'''\\
	\end{array}\right)\ ,
\end{equation}
where $\nu_0(x)$, $s_0(x)$, $s_1(x)$ and $s_2{x}$ are power series with leading term of order $x^{ak}$ and $\alpha,\alpha',\alpha'',\alpha'''\in\mathbf{C}$, also $m=ak$. The period $\tilde{\Pi}$ can be used to calculate the unipotent component of monodromy,
\begin{equation}
	\tilde{U}=\left(
\begin{array}{cccc}
 1 & 0 & 0 & 0 \\
 2 i \pi  & 1 & 0 & 0 \\
 -4 \pi ^2 & 4 i \pi  & 1 & 0 \\
 -8 i \pi ^3 & -12 \pi ^2 & 6 i \pi  & 1 \\
\end{array}
\right),\quad	\tilde{N}=\log(\tilde{U})=\left(
\begin{array}{cccc}
 0 & 0 & 0 & 0 \\
 2 i \pi  & 0 & 0 & 0 \\
 0 & 4 i \pi  & 0 & 0 \\
 0 & 0 & 6 i \pi  & 0 \\
\end{array}
\right)\ .
\end{equation}
And using the $\tilde{N}$ and $\tilde{\mathbf{a}}_m$, we can calculate
\begin{equation}
	\tilde{N}^3\tilde{\mathbf{a}}_m=\left(\begin{array}{c}0\\0\\0\\-48\pi^3 i \alpha\end{array}\right), \quad\text{ and }\quad
	\tilde{N}^4\tilde{\mathbf{a}}_m=0\ ,
\end{equation}
where the last part is expected since $\tilde{N}^4=0$ for $M$ point. Hence, for $M$ point, $d=3$ making it an infinite distance point.
\end{itemize}

\subsection{Geometric Realisation of the Hypergeometric Cases}
\label{sec:subgeoreal}
In this section, we will state the Calabi-Yau threefolds which are of interest to us. We start with a set of $n-3$ homogeneous polynomials of degree $d_1,...,d_{n-3}$ in the complex weighted projective space $\mathbb{P}^n(w_1,...,w_{n+1})$. A non-zero complete intersection, $Y$, of the zero locus of these polynomials will be a $3$ dimensional complex manifold. For this manifold to be Calabi-Yau, we need it's first Chern class to vanish, which translates to the following condition
\begin{equation}\label{calabicondtn}
	n+1=\sum_{k=1}^{n-3}d_k\ .
\end{equation}

In this following, we will study in detail the threefolds in projective space with all weights equal to one, i.e. $w_i=1$. It turns out that this subset in itself gives us a rich class of examples and has all the types of singular points mentioned above allowing us to draw general qualitative conclusions.

Since $d_k=1$ corresponds to a linear subspace of $\mathbb{P}^n$, which is $\mathbb{P}^{n-1}$, we restrict to $d_k>1$. This leads to only five possibilities:
\begin{itemize}
	\item One quintic in $\mathbb{P}^4$,
	\item one cubic and two quadratics in $\mathbb{P}^{6}$,
	\item one quartic and one quadratic in $\mathbb{P}^5$,
	\item two cubics in $\mathbb{P}^5$, and
	\item four quadratics in $\mathbb{P}^7$.
\end{itemize}
As mentioned before, we are going to work with the mirrors of these manifolds. Let's denote the mirror of a given complete intersection by $X_{d_1,...,d_{n-3}}(w_1,...,w_{n+1})$\footnote{We will use this notation when talking about a specific manifold. For general manifolds, we will use $X$.}. Hence, mirrors of the list above can be conveniently written as $X_{5}(1^5)$, $X_{3,2,2}(1^7)$, $X_{4,2}(1^6)$, $X_{3,3}(1^6)$ and $X_{2,2,2,2}(1^8)$\footnote{We will drop the weights in the notation in the following unless needed.}. The mirror of the quintic $X_5$ can be calculated by quotient action on the general quintic by the group $\mathbb{Z}_5^3$~\cite{GREENE199015}. For $X_{3,3}$, the mirror was correctly guessed in~\cite{1993alg.geom..1001L}. In~\cite{Berglund94}, the authors find the mirror for the rest of the cases. To parameterise the moduli space, we will switch between two coordinate systems depending on convenience. These are $z$ and $\psi$ related as 
\begin{equation}\label{eqn:zandpsi}
	z=\frac{1}{\mu\psi^{n+1}},\quad\text{for}\quad 0\leq\text{Arg}(z)<2\pi,\quad 0\leq\text{Arg}(\psi)<\frac{2\pi}{n+1}\ .
\end{equation}

\section{Calculation of Periods on $\mathcal{M}_{cs}(X)$}\label{sec:periods}
The period vector $\Pi$ is the solution of the following Picard-Fuchs equation~\cite{1993alg.geom..1001L,Hosono:1994ax}
\begin{equation}\label{eqn:sth_pfeqn}
    \mathcal{L}f_m(z)=\left(\theta^4-\mu z\prod_{k=1}^{4}(\theta+a_k)\right)f_m(z)=0\ ,
\end{equation}
where $\theta=z\frac{d}{dz}$ is the logarithmic derivative, the constants $\mu$ and $a_k$ are given in table~\ref{tab:allmetric2} and we define $\tilde{\Pi}(z)=\left(f_1(z),f_2(z),f_3(z),f_4(z)\right)^{\text{T}}$ as the period vector made up of four linearly independent solutions $f_m(z)$, $m=1,..,4$. The tilde, $\tilde{\ }$, in $\tilde{\Pi}(z)$ denotes that the period vector is not in a symplectic basis. The symplectic period vector $\Pi$ is most easily found at the $M$-point as suitable linear combinations of the solutions $f_i(z)$ as shown in the next section. 

The Riemann symbol for the Picard-Fuchs equation (\ref{eqn:sth_pfeqn})  summarises  the local exponents near its three singular points 
\begin{equation}\label{eqn:RSym}
\mathcal{P}
\left\{{\begin{array}{cccc} 0 & \frac{1}{\mu} & \infty &  \\ 0 & 0 & a_1 &  \\ 0 & 1 & a_2 & z \\ 0 & 2 & a_3 &  \\  0 & 1 & a_4 &  \\ \end{array}}\right\}\ 
.\end{equation}
From the discussion in section~\ref{sec:metricprop}, we see that our models have  a maximal unipotent $M$-point at $z=0$, a conifold  $C$-point  at $z=1/\mu$ and at $z=\infty$ either a $F$-,$C$-,$K$- or $M$-point which we will generically call the $s$-point. 
The complex structure moduli space is  $\mathcal{M}_{cs}(X)=\mathbb{P}^1\setminus\{z=0,z=1/\mu,z=\infty\}$. 
The periods $\Pi$ that we aim to calculate will be series solutions of the Picard-Fuchs equation expanded around different singular points 
and hence will have a finite radius of convergence. We label them as $\Pi_M$, $\Pi_C$ and $\Pi_s$ with radius of convergence $r_M$, $r_C$ and $r_s$ 
for the solutions around $M$-point, $C$-point and $s$-point respectively.  

Even though we analysed all cases in detail, we will use $X_{4,2}$ , $X_{3,3}$ and  $X_{2,2,2,2}$ as the key examples having an (orbifolded) 
$C$-point, $K$-point and $M$-point at $\psi=0$ respectively. After this rather general discussion about the periods, let's 
finally calculate them.
\subsection{Symplectic Basis at the $M$-Point and the Monodromy Group $\Gamma\subset {\rm Sp}(4,\mathbb{Z})$}
\label{sec:mpoint}
We start with results from~\cite{Hosono:1994ax,Klemm18}, where the authors calculate the periods in a  
symplectic basis around the $M$-point. This point corresponds  to a large radius point of the 
mirror family $Y$, so that the classical  intersection of $Y$  that occur in the $\hat \Gamma$-class, 
cf~\cite{Klemm18},  give sufficient  information to construct the  integral symplectic basis.       

One starts with defining
\begin{equation}\label{eqn:igamma} I_{\Gamma}(z,\epsilon)=\sum_{k=0}^{\infty}\frac{\prod_{l=1}^{n-3}\Gamma(d_l(k+\epsilon)+1)}{\prod_{l=1}^{n+1}\Gamma(w_l(k+\epsilon)+1)}z^{k+\epsilon}=\sum_{q=0}^{3}L_q(z)(2\pi i\epsilon)^q
\end{equation}
for a given threefold $X_{d_1,...,d_{n-3}}(w_1,...,w_{n+1})$. From this, $L_q(z)$ can be extracted by differentiating the equation~\eqref{eqn:igamma} $q$ times with respect to $\epsilon$ and then taking the limit $\epsilon\rightarrow 0$. The $L_q(z)$ constitutes the $\mathbb{Q}$\footnote{This means that the coefficients of the linear combination of $L_q(z)$, which gives us the period vector, are elements of the set $\mathbb{Q}$.} basis which combine to give us the period in symplectic basis as~\cite{Hosono:1994ax,Klemm18}
\begin{equation}\label{eqn:sth_pim}
    \Pi_M=\left({\begin{array}{c} F_0   \\ F_1 \\ X^0 \\ X^1 \end{array}}\right)=\left({\begin{array}{c} \int_{B_0}\Omega   \\ \int_{B_1}\Omega \\ \int_{A_0}\Omega \\ \int_{A_1}\Omega \end{array}}\right)=\left({\begin{array}{c} \kappa L_3+\frac{c_2\cdot D}{12} L_1   \\ -\kappa L_2+\sigma L_1 \\ L_0 \\ L_1 \end{array}}\right)\ ,
\end{equation}
where $\kappa$ and $c_2\cdot D$ are given in table~\ref{tab:allmetric2} and $\sigma=(\kappa\text{ mod }2)/2$. These periods have a radius of convergence $r_M=1/\mu$, which is nothing but the distance to the nearest singular point from  $M$-point. In the following, we will extend these solutions to the rest of the moduli space.
It follows immediately from the logarithmic terms that the monodromy induced by a counterclockwise loop 
around $z\; = \;0$ transforms the period as $\Pi \rightarrow M_{M} \Pi $ with
\begin{equation} 
M_{M}\; = \;\left(\begin{array}{cccc} 1& -1& \frac{\kappa}{6}+\frac{c_2\cdot D}{12} & \frac{\kappa}{2}+\sigma\\ 0& 1& \sigma-\frac{\kappa}{2}& -\kappa\\ 0& 0& 1& 0\\ 0& 0& 1& 1 \end{array}\right)\ . 
\label{maxmon}
\end{equation} 
The hypergeometric models have only three singular points, the above type $M$-point at $z=0$, 
a $C$-point, which is a normal conifold at $z=\mu^{-1}$ and at $1/z=0$ either an $F$-, $C$-, $K$- or $M$-point, 
all of them with an orbifold  action on top. Since w.r.t. to the symplectic   basis~\eqref{eqn:sth_pim} 
it is always the same cycle, namely the triple logarithmic one  that corresponds to the $D6$ brane on 
the mirror that vanishes, we can write an universal second monodromy  matrix for the conifold 
\begin{equation} 
M_{C}\; = \;\left(\begin{array}{cccc} 
1&0& 0& 0 \\ 
0&1& 0& 0\\ 
-1& 0& 1& 0\\ 
0& 0& 0& 1 \end{array}\right)\ . 
\label{Cmon}
\end{equation} 
Since ${\cal M}_{cs}=\mathbb{P}^1\setminus \{z=0,\mu,\infty\}$, we get for the third monodromy which distinguishes the 
$F,C,K,M$ type at $\frac{1}{z}=0$
\begin{equation} 
M_{1/z=0}\; = \; M_M M_C=\left(\begin{array}{cccc} 
1- \frac{\kappa}{6}-\frac{c_2\cdot D}{12}&-1&  \frac{\kappa}{6}+\frac{c_2\cdot D}{12}&  \frac{\kappa}{2}+\sigma\\ 
\frac{\kappa}{2}-\sigma&1&-\frac{\kappa}{2}+\sigma & -\kappa\\ 
-1& 0& 1& 0\\ 
-1& 0& 1& 1 \end{array}\right)\ . 
\label{Smon}
\end{equation} 
Any two of these three matrices generate the monodromy group $\Gamma\subset {\rm Sp}(4,\mathbb{Z})$.
%
\subsection{Period Degeneration near the  $C$-Point}\label{sec:cpoint}
To find periods around the point $\delta=1-\mu z$, we have to first solve the Picard-Fuchs equation and then find the matrix of transition to get periods in symplectic basis.

After change of variable from $z$ to $\delta$, equation~\eqref{eqn:sth_pfeqn} becomes
\begin{equation}
	\mathcal{L}f_m(\delta)=\left(x^4-(1-\delta)\prod_{k=1}^{4}(x+a_k)\right)f_m(\delta)=0\ ,
\end{equation}
where $x=(\delta-1)\frac{d}{d\delta}$. Using a series ansatz for $f_m(\delta)$, we find the following basis:
\begin{itemize}
	\item $X_{4,2}$:
	\begin{equation}\label{eqn:Pic}
	\tilde{\Pi}_C(\delta)=\left(
	\begin{array}{c}f_1(\delta)\\f_2(\delta)\\f_3(\delta)\\f_4(\delta)\\\end{array}\right)=\left(
	\begin{array}{c}
	 1+\frac{\delta^3}{256}+\frac{311 \delta^4}{49152}+\mathcal{O}(\delta^5) \\
	 \nu \\
	 \delta^2+\frac{119 \delta^3}{96}+\frac{23825 \delta^4}{18432}+\mathcal{O}(\delta^5) \\
	 \nu \log (\delta)-\frac{1045 \delta^3}{18432}-\frac{25013 \delta^4}{262144}+\mathcal{O}(\delta^5)\\
	\end{array}
	\right)\ ,
	\end{equation}
	where $\nu=\delta+\frac{23 \delta^2}{32}+\frac{1745 \delta^3}{3072}+\frac{31087 \delta^4}{65536}+\mathcal{O}(\delta^5)$ which is the unique vanishing period.
	\item $X_{3,3}$:
	\begin{equation}
	\tilde{\Pi}_C(\delta)=\left(
	\begin{array}{c}f_1(\delta)\\f_2(\delta)\\f_3(\delta)\\f_4(\delta)\\\end{array}\right)=\left(
	\begin{array}{c}
	 1+\frac{\delta ^3}{243}+\frac{175 \delta
	   ^4}{26244}+\mathcal{O}(\delta^5) \\
	  \nu \\
	 \delta ^2+\frac{67 \delta ^3}{54}+\frac{7549 \delta
	   ^4}{5832}+\mathcal{O}(\delta^5) \\
	 \nu  \log
	   (\delta )-\frac{323 \delta ^3}{5832}-\frac{58819 \delta
	   ^4}{629856}+\mathcal{O}(\delta^5) \\
	\end{array}
	\right)\ ,
	\end{equation}
	where $\nu=\delta +\frac{13 \delta ^2}{18}+\frac{139 \delta^3}{243}+\frac{12553 \delta ^4}{26244}+\mathcal{O}(\delta^5)$ is the unique vanishing period.
	\item $X_{2,2,2,2}$:
	\begin{equation}
	\tilde{\Pi}_C(\delta)=\left(
	\begin{array}{c}f_1(\delta)\\f_2(\delta)\\f_3(\delta)\\f_4(\delta)\\\end{array}\right)=\left(
	\begin{array}{c}
	 1+\frac{\delta ^3}{192}+\frac{13 \delta ^4}{1536}+\mathcal{O}(\delta^5) \\
	 \nu \\
	 \delta ^2+\frac{5 \delta ^3}{4}+\frac{377 \delta ^4}{288}+\mathcal{O}(\delta^5) \\
	 \nu\log (\delta)-\frac{13 \delta ^3}{288}-\frac{89 \delta ^4}{1152}+\mathcal{O}(\delta^5) \\
	\end{array}
	\right)\ ,
	\end{equation}
	where $\nu=\delta +\frac{3 \delta ^2}{4}+\frac{29 \delta ^3}{48}+\frac{49 \delta ^4}{96}+\mathcal{O}(\delta^5)$ is the unique vanishing period.
\end{itemize}

\begin{figure}[t]
        \centering
    \begin{tikzpicture}[scale=1.2]
        \draw[-] (0,-1.2*2) -- (0,3);
        \node [above] at (0,3) {Im$(z)$};
        \draw[-] (-1.2*2,0) -- (7,0);
        \node [right] at (7,0) {Re$(z)$}; 
        
        \draw[red,thick] (1*2,0) circle [radius=1*2];
        \draw[fill=red,opacity=0.3] (1*2,0) circle [radius=1*2];
        \draw[blue,thick] (0,0) circle [radius=1*2];
        \draw[fill=blue,opacity=0.3] (0,0) circle [radius=1*2]; 
        
        \draw[blue,fill] (0,0) circle [radius=0.05];
        \draw[red,fill] (1*2,0) circle [radius=0.05];
        \node [above right] at (1*2,0) {\footnotesize$1/\mu$};
        \draw[black,fill] (6.5,3) circle [radius=0.05];
        \node [below] at (6.5,3) {\footnotesize$z=\infty$};
        \node [above left] at (0,0) {\footnotesize$0$};
        \draw[black,fill] (1,0) circle [radius=0.05];
        \node [above] at (1,0) {\footnotesize$0.5/\mu$};
        
        \draw (1-0.25,2.5)--(5.85-0.25,2.5)--(5.85-0.25,3.5)--(1-0.25,3.5)--(1-0.25,2.5);
        \draw[blue,fill,opacity=0.3] (1.2-0.25,2.5+2/3+0.05) circle [radius=0.07];
        \draw[red,fill,opacity=0.3] (1.2-0.25,2.5+1/3-0.05) circle [radius=0.07];
        \node[right] at (1.2+0.05-0.25,2.5+2/3+0.05) {$\Pi_M$ convergence region};
        \node[right] at (1.2+0.05-0.25,2.5+1/3-0.05) {$\Pi_{C}$ or $\tilde{\Pi}_C$ convergence region};
        \normalsize

    \end{tikzpicture}
\caption{The convergence region of periods expanded around  $M$- and  $C$- point in $\mathcal{M}_{cs}$.}
\label{fig:MandC}
\end{figure}
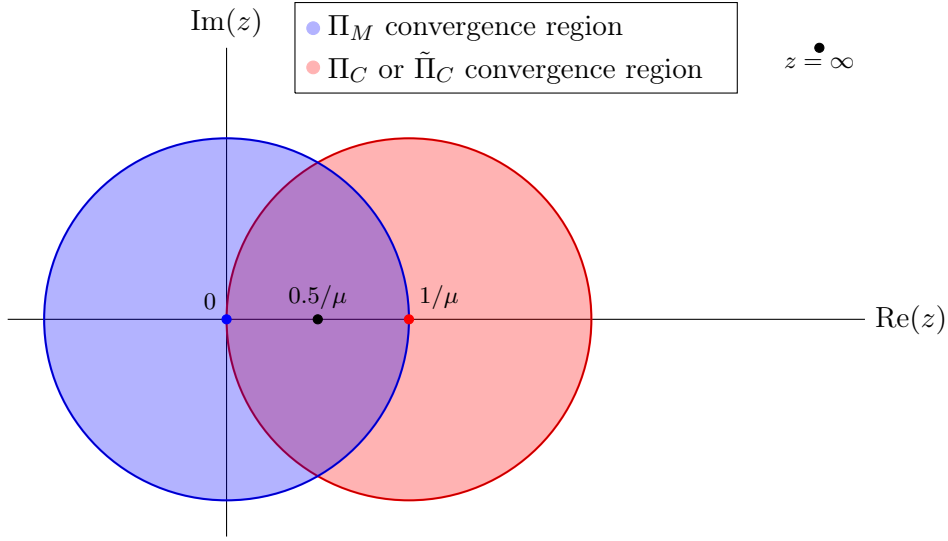

The radius of convergence of the periods calculated above is $r_C=1/\mu$\footnote{This is the convergence radius in $z$ coordinates. In $\delta$ coordinates, $r_C=1$.}, which is again the distance to the nearest singular point from the $C$-point. The periods $\tilde{\Pi}_C$ calculated above are in an arbitrary basis, but we want periods in symplectic basis. Hence, we want to find the continuation matrix $T_{MC}$ satisfying
\begin{equation}
	\Pi_{C}(\delta)=T_{MC}\tilde{\Pi}_{C}(\delta)\ .
\end{equation}
In figure~\ref{fig:MandC}, we show the convergence region of the periods we have calculated. Note that $\Pi_M$ and $\Pi_C$ are both convergent in the region common to blue and red. Hence, select the point in the middle, $z=0.5/\mu$ (or $\delta=0.5$), and at this point $\Pi_M=\Pi_C$ therefore we have
\begin{equation}
	\Pi_{C}(\delta=0.5)=T_{MC}\tilde{\Pi}_{C}(\delta=0.5)=\Pi_M(z=0.5/\mu)\ .
\end{equation}
In the equation above, we know $\tilde{\Pi}_C$ and $\Pi_M$, and hence, we can solve for $T_{MC}$ by expanding the solutions around $z=1/\mu$ and comparing the coefficients. Following are the $T_{MC}$ we find:
\begin{itemize}
	\item $X_{4,2}$:
	\begin{equation}\label{eqn:tmc1}
	T_{MC}=\left(
	\begin{array}{cccc}
	 0 &  -0.45016 i & 0 & 0 \\
	 7.2218 & 1.4000 & -0.20681 & 0 \\
	 1.0873 & -0.028048 & 0.006033 & 0.071645 \\
	 1.1186 i & +0.14909 i & -0.029801 i & 0 \\
	\end{array}
	\right)\ 
	,\end{equation}
	\item $X_{3,3}$:
	\begin{equation}\label{eqn:tmc2}
	T_{MC}={\footnotesize\left(
	\begin{array}{cccc}
	 0 &  -0.47746 i & 0 & 0 \\
	 7.2268\, +0.53267 i & 1.4974\, +0.074299 i &
	   -0.22102-0.015039 i & 0 \\
	 1.0922 & -0.029110 & 0.006133 & 0.075991 \\
	 1.0653 i & 0.14859 i & -0.030072 i & 0
	   \\
	\end{array}
	\right)}\ ,
	\end{equation}
	\item $X_{2,2,2,2}$:
	\begin{equation}\label{eqn:tmc3}
	T_{MC}=\left(
	\begin{array}{cccc}
	 0 & -0.6366 i & 0 & 0 \\
	 8.9491 & 2.2373 & -0.3622 & 0 \\
	 1.1186 & -0.032354 & 0.0053906 & 0.10132 \\
	 0.90273 i & 0.14610 i & -0.032929 i & 0 \\
	\end{array}
	\right)\ .
	\end{equation}
\end{itemize}

\subsection{Around  $s$-Point ($\psi= 0$)}\label{sec:spoint}
Finally, we will find the converging periods around the  $s$-point at $w=1/z=0$\footnote{To increase confusion, we have three variables now: $z$, $w$ and $\psi$ and we will use them interchangeably whenever convenient.}. First, just like in section~\ref{sec:cpoint}, we rewrite equation~\eqref{eqn:sth_pfeqn} after change of variables from $z$ to $w$, getting
\begin{equation}\label{eqn:PFw}
\mathcal{L}f_m(w)=\left(\vartheta^4-\mu \frac{1}{w} \prod_{k=1}^4(\vartheta-a_k)\right)f_m(w)=0\ ,
\end{equation}
where $\vartheta=w\frac{d}{dw}$. This equation can be solved using Meijer G-functions and the complete calculation is shown in appendix~\ref{app:mgfunction}. Here, we will just state the resulting functions.

\begin{itemize}
	\item $X_{4,2}$:
	\begin{equation}
        \begin{array}{ll}
        f_1(w)=G_{4,4}^{1,4}\left(-\frac{w}{2^{10}}|
        \begin{array}{c}
         1,1,1,1 \\
         \frac{1}{4},\frac{1}{2},\frac{1}{2},\frac{3}{4} \\
        \end{array}
        \right), &\quad f_2(w)=G_{4,4}^{1,4}\left(-\frac{w}{2^{10}}|
        \begin{array}{c}
         1,1,1,1 \\
         \frac{1}{2},\frac{1}{4},\frac{1}{2},\frac{3}{4} \\
        \end{array}
        \right),\\
		  \null\\
        f_3(w)=G_{4,4}^{1,4}\left(-\frac{w}{2^{10}}|
        \begin{array}{c}
         1,1,1,1 \\
         \frac{3}{4},\frac{1}{4},\frac{1}{2},\frac{1}{2} \\
        \end{array}
        \right)\text{ and} &\quad f_4(w)=G_{4,4}^{2,4}\left(\frac{w}{2^{10}}|
        \begin{array}{c}
        1,1,1,1 \\
        \frac{1}{2},\frac{1}{2},\frac{1}{4},\frac{3}{4} \\
        \end{array}
        \right)\ .
        \end{array}
	\end{equation}
	\item $X_{3,3}$:
	\begin{equation}\label{eqn:sX33}
         \begin{array}{ll}
             f_1(w)=G_{4,4}^{1,4}\left(-\frac{w}{3^6}|
             \begin{array}{c}
              1,1,1,1 \\
              \frac{1}{3},\frac{1}{3},\frac{2}{3},\frac{2}{3} \\
             \end{array}
             \right), &\quad f_2(w)=G_{4,4}^{1,4}\left(-\frac{w}{3^6}|
             \begin{array}{c}
              1,1,1,1 \\
              \frac{2}{3},\frac{1}{3},\frac{1}{3},\frac{2}{3} \\
             \end{array}
             \right),\\
			 	\null\\
             f_3(w)=G_{4,4}^{2,4}\left(\frac{w}{3^6}|
             \begin{array}{c}
              1,1,1,1 \\
              \frac{1}{3},\frac{1}{3},\frac{2}{3},\frac{2}{3} \\
             \end{array}
             \right),\text{ and} &\quad f_4(w)=G_{4,4}^{2,4}\left(\frac{w}{3^6}|
             \begin{array}{c}
              1,1,1,1 \\
              \frac{2}{3},\frac{2}{3},\frac{1}{3},\frac{1}{3} \\
             \end{array}
             \right)\ .
         \end{array}
	\end{equation}
	\item $X_{2,2,2,2}$:
	\begin{equation}
	    \begin{array}{ll}
	    f_1(w)=G_{4,4}^{1,4}\left(-\frac{w}{2^8}|
	    \begin{array}{c}
	     1,1,1,1 \\
	     \frac{1}{2},\frac{1}{2},\frac{1}{2},\frac{1}{2} \\
	    \end{array}\right), &\quad f_2(w)=G_{4,4}^{2,4}\left(\frac{w}{2^8}|
	    \begin{array}{c}
	     1,1,1,1 \\
	     \frac{1}{2},\frac{1}{2},\frac{1}{2},\frac{1}{2} \\
	    \end{array}
	    \right),\\
	    \null\\
	    f_3(w)=G_{4,4}^{3,4}\left(-\frac{w}{2^8}|
	    \begin{array}{c}
	     1,1,1,1 \\
	     \frac{1}{2},\frac{1}{2},\frac{1}{2},\frac{1}{2} \\
	    \end{array}
	    \right),\text{ and}& \quad f_4(w)=G_{4,4}^{4,4}\left(\frac{w}{2^8}|
		\begin{array}{c}
	     1,1,1,1 \\
	     \frac{1}{2},\frac{1}{2},\frac{1}{2},\frac{1}{2} \\
	    \end{array}
	    \right)\ .
	    \end{array}
	\end{equation}
\end{itemize}

These functions have a Mellin-Barnes type integral representation which allows us to find converging solutions in the region $|w|<\mu$ and $|w|>\mu$ by changing the contour (see appendix~\ref{app:mgfunction}). This provides us with an analytic continuation of the period $\tilde{\Pi}_s(w)=\left(f_1(w),f_2(w),f_3(w),f_4(w)\right)^{\text{T}}$ from $|w|<\mu$ to $|w|>\mu$. Using the form of $\tilde{\Pi}_s(w)$ in $|w|>\mu$ (or $|z|<1/\mu$), and matching first few coefficients with $\Pi_M(z)$, we can find the transition matrix $T_{Ms}$ satisfying
\begin{equation}
	\Pi_M(z)=T_{Ms}\tilde{\Pi}_s(w)\ .
\end{equation}
The resulting matrices are:
\begin{itemize}
	\item $X_{4,2}$:
	\begin{equation}
	T_{Ms}=\left(
	\renewcommand{\arraystretch}{1.4}
	\begin{array}{cccc}
	 \frac{{1}+{i}}{2\pi ^2} & 0 & -\frac{{1}-{i}}{2\pi ^2} & -\frac{2 i \sqrt{2}}{\pi ^3} \\
	 \frac{1-i}{\pi ^2} & \frac{2 i \sqrt{2}}{\pi ^2} & -\frac{1+i}{\pi ^2} & 0 \\
	 \frac{1}{2 \pi ^2} & \frac{i \sqrt{2}}{\pi ^2} & -\frac{1}{2 \pi ^2} & -\frac{i \sqrt{2}}{\pi ^3} \\
	 \frac{1+i}{4\pi ^2} & 0 & -\frac{1-i}{4\pi ^2} & -\frac{i}{\sqrt{2} \pi ^3} \\
	\end{array}
	\right)\ 
	,\end{equation}
	\item $X_{3,3}$:
	\begin{equation}
	T_{Ms}=\left(
	\renewcommand{\arraystretch}{1.4}
	\begin{array}{cccc}
	 -\frac{9 \left(3+i \sqrt{3}\right)}{8 \pi ^2} & \frac{9 \left(3-i \sqrt{3}\right)}{8 \pi ^2} & \frac{9 i}{8 \pi ^3} & \frac{9 i}{8 \pi ^3} \\
	 -\frac{3+i \sqrt{3}}{8 \pi ^2} & \frac{3-i \sqrt{3}}{8 \pi ^2} & \frac{21 \left(1-i \sqrt{3}\right)}{8 \left(\sqrt{3}-5 i\right) \pi ^3} & \frac{3 i \left(4
	   \sqrt{3}+i\right)}{4 \left(\sqrt{3}-5 i\right) \pi ^3} \\
	 -\frac{3 \left(5+i \sqrt{3}\right)}{8 \pi ^2} & \frac{3 \left(5-i \sqrt{3}\right)}{8 \pi ^2} & \frac{3 \left(\sqrt{3}+3 i\right)}{16 \pi ^3} & -\frac{3 \left(\sqrt{3}-3
	   i\right)}{16 \pi ^3} \\
	 -\frac{3+i \sqrt{3}}{4 \pi ^2} & \frac{3-i \sqrt{3}}{4 \pi ^2} & \frac{3 i}{8 \pi ^3} & \frac{3 i}{8 \pi ^3} \\
	\end{array}
	\right)\ 
	,\end{equation}
	\item $X_{2,2,2,2}$:
	\begin{equation}
	    T_{Ms}=\left(
	\renewcommand{\arraystretch}{1.4}
	\begin{array}{cccc}
	 0 & -\frac{4 i}{\pi ^3} & 0 & \frac{2 i}{\pi ^5} \\
	 0 & \frac{4 i}{\pi ^3} & -\frac{4 i}{\pi ^4} & 0 \\
	 \frac{i}{\pi^2} & -\frac{i}{\pi ^3} & -\frac{i}{\pi ^4} & \frac{i}{\pi ^5} \\
	 0 & -\frac{i}{2 \pi ^3} & 0 & \frac{i}{2 \pi ^5} \\
	\end{array}
	\right)\ 
	.\end{equation}
\end{itemize}

Using the $T_{Ms}$, we can find periods in symplectic basis for $|z|>1/\mu$ using
\begin{equation}
	\Pi_s(w)=T_{Ms}\tilde{\Pi}_s(w)\ .
\end{equation}
\section{Global Structure of the Metric on $\mathcal{M}_{cs}(X)$}\label{sec:metric}
With the periods calculated on the whole moduli space $\mathcal{M}_{cs}$, we can now find the K\"{a}hler potential using~\eqref{eqn:Kahler2} and then find the metric and its scalar curvature using equation~\eqref{eqn:Metric} and equation~\eqref{eqn:scalar} respectively. In the following, we will plot the resulting metric for each threefold's complex structure moduli space and talk qualitatively about the distances to the singular points. These distances are measured from any non-singular point on the threefold.

\begin{itemize}
	\item $X_{4,2}$: The metric for this threefold is shown in figure~\ref{fig:metric}, and table~\ref{table:glimit} shows the limiting behaviour of the metric and the scalar curvature as we approach the singular points. From the limiting behaviour, we can see that there is only one point at infinite distance, the  $M$-point at $\psi=\infty$. The point $\psi=0$ is at finite distance. At the conifold point, $\psi=1$, the metric and curvature diverges but the distance is finite.
	\begin{figure}[t]
	    \centering
	\begin{overpic}[scale=0.35]{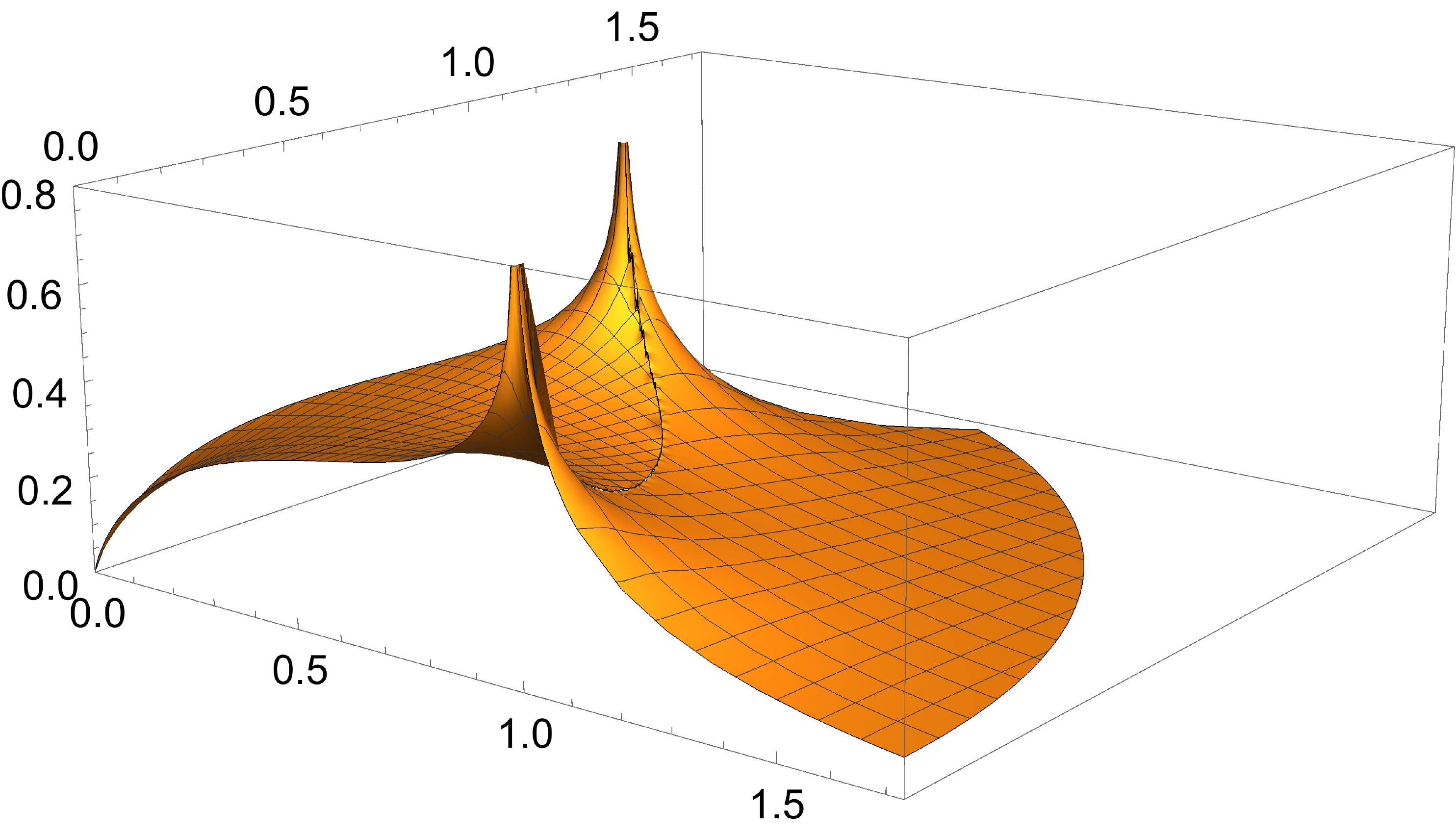}
	 \put (-8,30) {$G_{\psi\overline{\psi}}$}
	 \put (24,4) {Re$(\psi)$}
	 \put (19,54) {Im$(\psi)$}
	\end{overpic}
	 \caption{The metric $G_{\psi\overline{\psi}}$ plotted as function of $\psi$ for $\mathcal{M}_{cs}(X_{4,2})$.}
	 \label{fig:metric}
	\end{figure}
	\begin{table}[t]
	\centering
	\begin{tabular}{|c|>{\centering}m{3.5cm}|c|}
	\hline
	$\psi$ & $G_{\psi\overline{\psi}}$\vspace{1mm} & $R$  \\ \hline
	  0 &  \vspace{1mm}$-\frac{27 \pi ^2 \Gamma \left(-\frac{1}{4}\right)^4 \left| \psi\right|  \log (\left| \psi \right| )}{8 \Gamma\left(\frac{1}{4}\right)^8}$\vspace{1mm} & $-\frac{4 \Gamma \left(\frac{1}{4}\right)^8}{27 \pi ^2 \Gamma \left(-\frac{1}{4}\right)^4 \left|\psi\right|^3 \log^3(\left|\psi\right|))}$\\ \hline
	  1 &   \vspace{1mm}$-\frac{6^2 \kappa  \log (r)}{(2\pi)^3 \text{a g}}$\vspace{1mm}  &  $-\frac{ \text{a g }(2\pi)^3}{2\times6^2 \kappa\ r^2 \log^3(r)}$ \\ \hline
	  $\infty$ &  \vspace{1mm}$\frac{3}{4 \left| \psi \right| ^2 \log ^2(\left| \psi \right|)}$\vspace{1mm}  & $-\frac{4}{3}$ \\ \hline
	\end{tabular}
	\caption{The behaviour of $G_{\psi\overline{\psi}}$ as $\psi$ approaches the singular points for $\mathcal{M}_{cs}(X_{4,2})$. Here $r=\left|\psi-1\right|$, a is real part of $(T_{MC})_{2,1}$, g is imaginary part of $(T_{MC})_{4,1}$, the matrix $T_{MC}$ is given in equation~\eqref{eqn:tmc1} and $\kappa=8$ is given in table~\ref{tab:allmetric2}.}
	\label{table:glimit}
	\end{table}
	
	\item $X_{3,3}$: For this case, figure~\ref{fig:metric2} shows the metric and table~\ref{table:glimit2} shows limiting behaviour of the metric and the scalar curvature. Similar to $X_{4,2}$ discussed above, for this threefold  $M$-point is at infinite distance and the conifold point is finitely away from any non-singular point. Here, a new feature is observed at the  $s$-point, it is at infinite distance.
	\begin{figure}[t]
		\centering
	\begin{overpic}[scale=0.35]{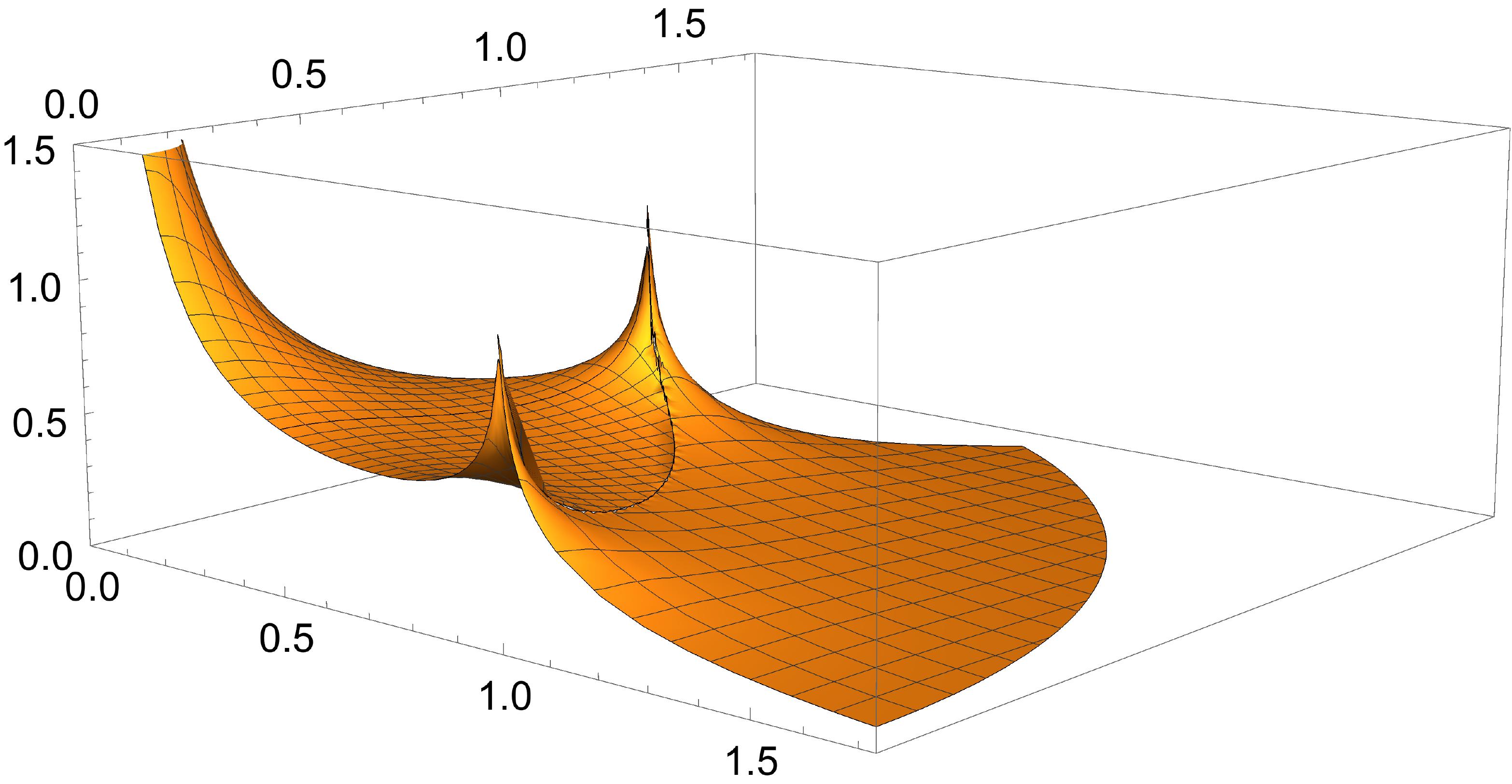}
	 \put (-7,28) {$G_{\psi\overline{\psi}}$}
	 \put (22,3) {Re$(\psi)$}
	 \put (19,51) {Im$(\psi)$}
	\end{overpic}
	 \caption{The metric $G_{\psi\overline{\psi}}$ plotted as function of $\psi$ for $\mathcal{M}_{cs}(X_{3,3})$.}
	 \label{fig:metric2}
	\end{figure}
	\begin{table}[t]
	\centering
	\begin{tabular}{|c|>{\centering}m{3cm}|c|}
	\hline
	$\psi$ & $G_{\psi\overline{\psi}}$\vspace{1mm} & $R$  \\ \hline
	  0 & \vspace{1mm}$\frac{1}{4 \left| \psi \right| ^2 \log ^2(\left| \psi \right| )}$\vspace{1mm} & $-4$ \\ \hline
	  1 &   \vspace{1mm}$-\frac{6^2 \kappa  \log (r)}{(2\pi)^3 \text{a g}}$\vspace{1mm} &  $-\frac{ \text{a g }(2\pi)^3}{2\times 6^2 \kappa\ r^2 \log^3(r)}$ \\ \hline
	  $\infty$ &  \vspace{1mm}$\frac{3}{4 \left| \psi \right| ^2 \log ^2(\left| \psi \right|
	   )}$\vspace{1mm}& $-\frac{4}{3}$ \\ \hline
	\end{tabular}
	\caption{The behaviour of $G_{\psi\overline{\psi}}$ as $\psi$ approaches the singular points for $\mathcal{M}_{cs}(X_{3,3})$. Here $r=\left|\psi-1\right|$, a is real part of $(T_{MC})_{2,1}$, g is imaginary part of $(T_{MC})_{4,1}$, the matrix $T_{MC}$ is given in equation~\eqref{eqn:tmc2} and $\kappa=9$ is given in table~\ref{tab:allmetric2}.}
	\label{table:glimit2}
	\end{table}
	
	\item $X_{2,2,2,2}$: For this case, figure~\ref{fig:metric3} shows the metric and table~\ref{table:glimit3} has information about limiting behaviour of the metric and the scalar curvature. The metric for this threefold is qualitatively similar to that of $X_{3,3}$ with a more severe singularity at $\psi=0$.
	\begin{figure}[t]
		\centering
	\begin{overpic}[scale=0.33]{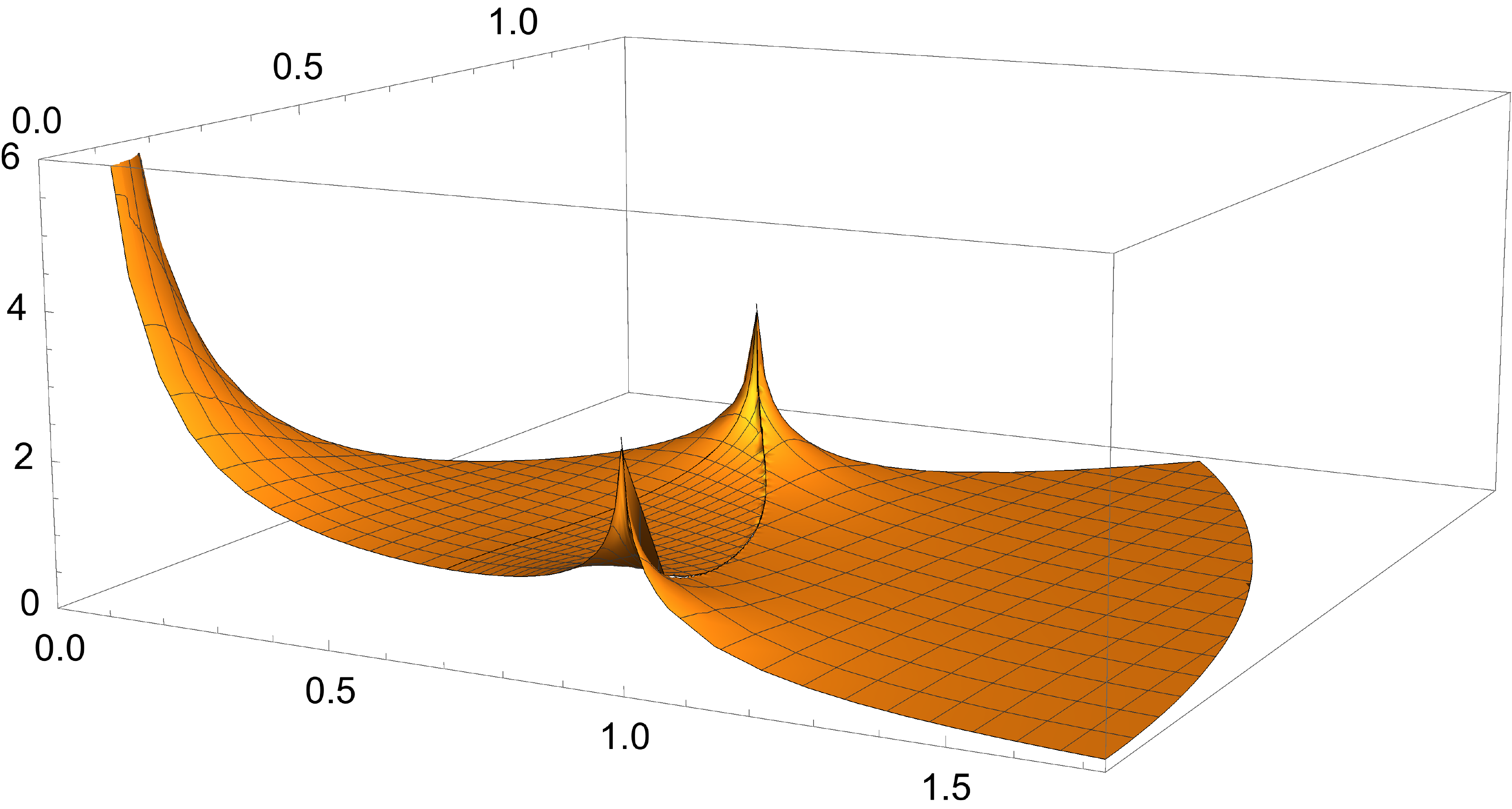}
	 \put (-7,27) {$G_{\psi\overline{\psi}}$}
	 \put (28,3) {Re$(\psi)$}
	 \put (15,52) {Im$(\psi)$}
	\end{overpic}
	 \caption{The metric $G_{\psi\overline{\psi}}$ plotted as function of $\psi$ for $\mathcal{M}_{cs}(X_{2,2,2,2})$.}
	 \label{fig:metric3} 
	\end{figure}
	\begin{table}[t]
	\centering
	\begin{tabular}{|c|>{\centering}m{3cm}|c|}
	\hline
	$\psi$ & $G_{\psi\overline{\psi}}$\vspace{1mm} & $R$  \\ \hline
	  0 & \vspace{1mm}$\frac{3}{4 \left| \psi \right| ^2 \log ^2(\left| \psi \right| )}$\vspace{1mm} & $-\frac{4}{3}$ \\ \hline
	  1 &   \vspace{1mm}$-\frac{8^2 \kappa  \log (r)}{(2\pi)^3 \text{a g}}$\vspace{1mm} &  $-\frac{ \text{a g }(2\pi)^3}{2\times 8^2 \kappa\ r^2 \log^3(r)}$ \\ \hline
	  $\infty$ &  \vspace{1mm}$\frac{3}{4 \left| \psi \right| ^2 \log ^2(\left| \psi \right|
	   )}$\vspace{1mm}& $-\frac{4}{3}$ \\ \hline
	\end{tabular}
	\caption{The behaviour of $G_{\psi\overline{\psi}}$ as $\psi$ approaches the singular points for $\mathcal{M}_{cs}(X_{2,2,2,2})$. Here $r=\left|\psi-1\right|$, a is real part of $(T_{MC})_{2,1}$, g is imaginary part of $(T_{MC})_{4,1}$, the matrix $T_{MC}$ is given in equation~\eqref{eqn:tmc3} and $\kappa=16$ is given in table~\ref{tab:allmetric2}.}
	\label{table:glimit3}
	\end{table}
\end{itemize}

To summarise, we have two points of special interest: $\psi=\infty$ which is at infinite distance for every threefold, and $\psi=0$ which is at infinite distance for $X_{3,3}$ and $X_{2,2,2,2}$. If SDC holds, we expect to find a tower of exponentially light states as we approach these points. We will discuss each of the two points in the following.

\section{Swampland Distance Conjecture}\label{sec:sdc}
In the following, we will look at the two infinite distance points discussed above. We will identify the 
light states with  $D2-D0$ bound states at the  $M$-point and since we don't have a type  IIA interpretation yet, at the $K$-point, we will identify the light states 
with BPS 3-branes.

\subsection{Light States near the $M$-Point}\label{sec:psi_infty}

\begin{figure}[t]
    \centering
\begin{overpic}[scale=0.3]{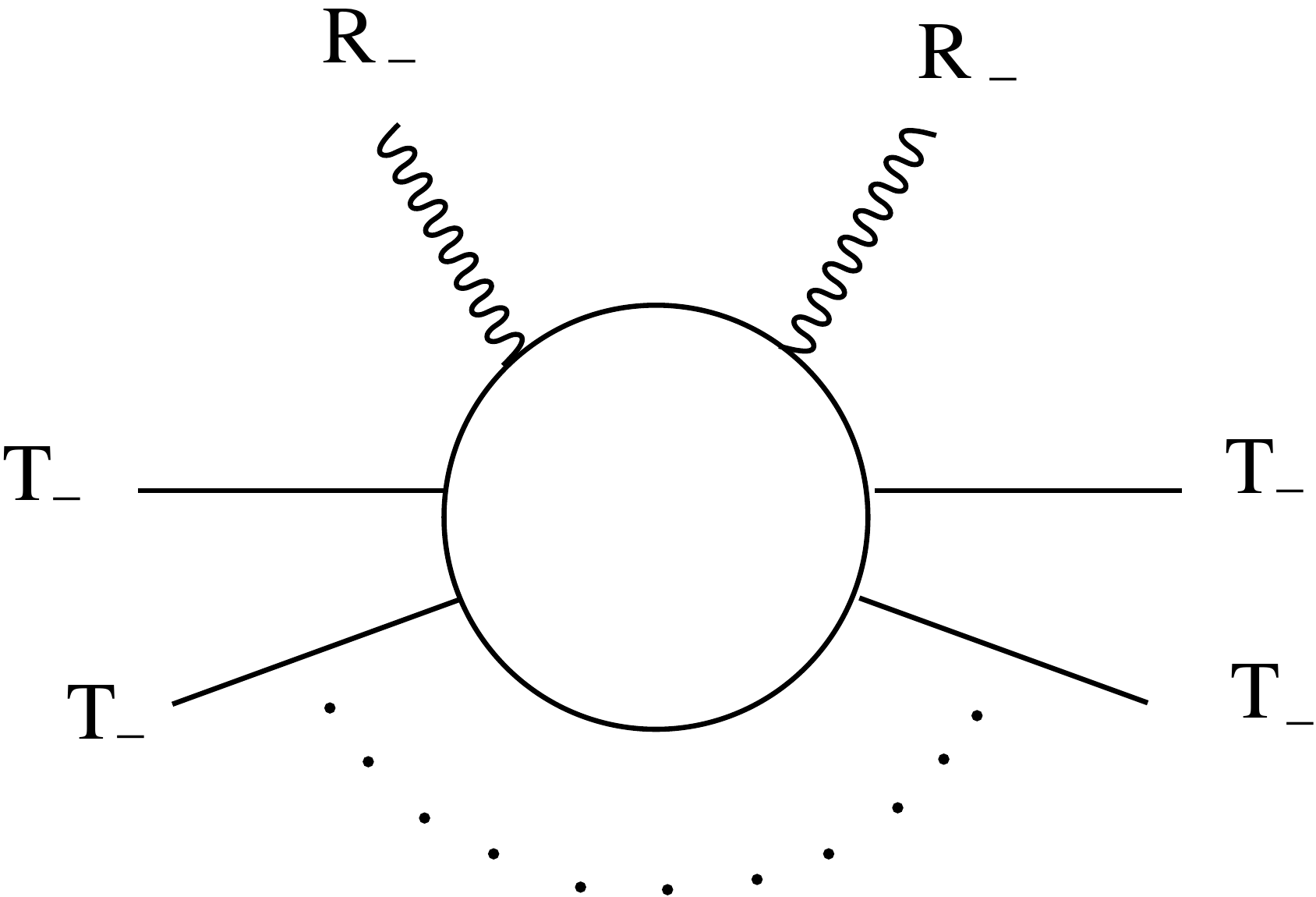}
\end{overpic}
 \caption{One loop integral counting the BPS index $I^\beta_g$ with BPS states running in the loop.}
 \label{fig:schwingerloop}
\end{figure}

Light BPS states can be identified by their contribution to the Schwinger one loop amplitude creating the coupling of the 
anti selfdual part of the curvature   $R_-$ to the  anti selfdual part of the graviphoton 
field strength  $T_-$.  This  term resides in the vector multiplet sector of  the effective 4d 
$N=2$ action and its  dependence on the K\"ahler moduli $t$ is given  by the higher genus topological string amplitudes $\mathcal{F}_g(t)$ as 
\begin{equation}
\sum_{g>0} \int_{M_{1,3}} d x^4 \mathcal{F}_g(t) T_-^{2g-2} R_-^2 \ . 
\end{equation}
As pointed out  in~\cite{Antoniadis:1993ze,Gopakumar:1998jq},  this moduli dependence can be   calculated 
by the Schwinger one-loop integral shown in figure~\ref{fig:schwingerloop},
where  the topological string coupling $\lambda$ becomes formally identified with the $T_-$ 
insertions.  

Generally the BPS states contribute to the Schwinger one loop  amplitude only via a BPS index.  In the large volume  
limit the index $I^\beta_g$ is directly related to the BPS degeneracies $N^\beta_{j_Lj_R}$ 
by the following re-organisation of the spin representations $[j_L]_L,[j_R]_L$ of the $5$d little group 
${\rm SU}(2)_L\times {\rm SU}(2)_R$  
\begin{equation}
\sum_{J_L j_R\in \frac{1}{2} \mathbb{N}} (-1)^{j_R}( 2J_+ + 1) N^\beta_{j_Lj_R} [j_L]_L=\sum_{g=0}^\infty I^\beta _g \left(2[0]_L+\left[\frac{1}{2}\right]\right)^{\otimes g}\ .
\end{equation}
As explained  in~\cite{Gopakumar:1998jq,Katz:1999xq} and later (mathematically) rigorously 
in~\cite{MR2552254} using stable pair invariants, these BPS  invariants can be viewed in the large 
volume limit as bound states of $D2$ branes interpreted as pure sheaf ${\cal F}$ of complex dimension one carrying the charge ${\rm ch}_2({\cal F})
=\beta\in H_2(M,\mathbb{Z})$ and $D0$ branes  whose charge is related to $n=\chi({\cal F})\in \mathbb{Z}$ and becomes 
after some one-to-one mapping identified with the genus of the topological string amplitude.         

The result of the combined analysis in~\cite{Antoniadis:1993ze,Gopakumar:1998jq} 
is that the all genus free energy of the  topological string can be written as  
\begin{equation}
\begin{array}{rcl}
{\cal F}(\lambda,t)&=&\displaystyle{\sum_{g=0}^\infty \lambda^{2 g-2}  \mathcal{F}_g(t)}\\
&=&\displaystyle{\frac{c(t)}{\lambda^2}+l(t)+\sum_{g=0}^\infty \sum_{\beta\in H_2(M,\mathbb{Z})}\sum_{m=1}^\infty I_g^\beta \frac{1}{m}
\left(2 \sin \frac{m \lambda}{2}\right)^{2 g-2}  Q^{\beta m}}\ ,\\
\end{array}
\label{schwingerloop}
\end{equation}
with $Q^{\beta}=e^{2 \pi i t}$, with $t$ unique  K\"ahler parameter for the one-parameter cases. Here $\mathcal{F}_0$, the genus zero contribution, 
is the prepotential, calculated using~\eqref{eqn:sth_pim}. However the classical terms, i.e.  the inhomogeneous cubic polynomial 
$c(t)$ in $\mathcal{F}_0$ and the linear term $l(t)$ in $\mathcal{F}_1$ in $t\sim\log(z)$ are not calculated by the Schwinger loop contribution 
in~\cite{Gopakumar:1998jq}\footnote{Except for the constant term  $\frac{\chi \zeta(3)}{2 (2 \pi i)^3}$ in $c(t)$, which comes from
pure $D0$ brane charges and gives only a sub-leading contribution to the metric.}. 

Linear logarithmic  contributions to $\mathcal{F}_0$ have been calculated from the one loop wave function regularisation contributing 
to the gauge coupling of one photon\footnote{This photon is the one that couples to the light charges state.} in the  generic 
$U(1)^{h_{11}}$ gauge group in the vector moduli space of Calabi-Yau compactifications, i.e. in $N=2$ supergravity, 
in~\cite{Strominger:1995cz} at the conifold point analogously  as in $N=2$  gauge theory in~\cite{Seiberg:1994rs} at the 
monopole point. In both cases, there is only one particle namely the magnetic monopole or the massless extremal black 
hole running  the loop and the calculation is simple. 

In the context of the SDC it would be satisfying to show that the integrating out the infinite tower of states yields the structure of 
the  higher logarithmic terms in $\mathcal{F}_0$ at the $M$-point (and the $K$- points) that give rise to the infinite metric distance property. 
While we have not attempted this calculation in  this  project, it is likely that it will work only with the particular spectrum encoded 
in~\eqref{schwingerloop}  as the asymptotic  growth of these states knows about the classical intersection numbers  that are 
the coefficients of the cubic in $c(t)$ as it is predicted from the microscopic black hole entropy counting arguments and was 
checked  in~\cite{Huang:2007sb}. 
 
While the result (\ref{schwingerloop}) is specific for the $D2-D0$ brane spectrum at large radius points, 
the evaluation of the Schwinger one-loop diagram is possible whenever the BPS spectrum is known 
at a singular point and also the topological string amplitudes can be calculated at other  singular points 
in the B-model~\cite{Huang:2006hq}. For example the contribution of the light particle at the conifold  
to the $\mathcal{F}_{g>0}(t_c)$ amplitudes in the leading order in the flat coordinate $t_c$ at the conifold  
has been evaluated via the Schwinger loop calculation  and gives rise to the conifold gap~\cite{Huang:2006hq}. 
The logarithmic contribution to $\mathcal{F}_1(t_c)= \frac{1}{12} t_c+\ldots$ at the conifold has been obtained  
previously in~\cite{Vafa:1995ta}. Also, one can perform calculation to find behaviour of the state which becomes massless at $C$-point. For the exponents $(a,b,b,c)$, near the singularity, one finds
\begin{equation}
	m\sim \alpha|\tilde{x}|^{b-a}\ ,\quad\quad d(\tilde{x},r_0)=-\beta\int_{r_0}^{|\tilde{x}|}dx\left(-x^{-2a}\log(x)\right)^{1/2}\ ,
\end{equation} 
where $\tilde{x}$ is the local coordinate such that singular point is at $\tilde{x}=0$, $r_0>\tilde{x}$, $d(\tilde{x},r_0)$ is the distance between $\tilde{x}$ and $r_0$, and $\alpha,\beta\in\mathbb{R}^{+}$. Clearly, mass vanishes at $\tilde{x}=0$, and distance to singularity, i.e. $d(0,r_0)$, is finite since $a<1/2$. One can find the coordinate $\tilde{x}$ in terms of distance $d(\tilde{x},r_0)$ by inverting the second function above.

In any case at the $M$- points the $D2$-$D0$ bound states in the type IIA compactification are the desired 
infinite tower of stable states, whose mass of the constituents vanishes exponentially with the distance as one approaches the $M$-point,
\begin{equation}
m_0^p\sim e^{-\sqrt{3}d(p,q)}m_0^q\ , \quad\quad m_2^p\sim e^{-\frac{1}{\sqrt{3}}d(p,q)}m_2^{q}
\end{equation}
and is dominated by the $D2$-brane mass. This result is calculated by performing analogous analysis to that presented in detail in section~\ref{sec:psi_0}.  Moreover in the topological sector, mirror symmetry applies 
and yields the very same prediction for the type IIB light BPS states.
  
Let us compare this behaviour  with the one of Kaluza Klein states proposed in~\cite{Blumenhagen:2018nts}. 
On performing dimensional reduction, one finds that the mass of these KK states scales as
\begin{equation}\label{eqn:chap2KK}
	m_{KK}^2\sim\frac{1}{R^2}\ ,
\end{equation}
where $R^2$ is the size of the manifold. For the threefolds considered here, in the large volume limit, $R^2$ is given by the imaginary part of the K\"{a}hler modulus (area), $t$, on $\mathcal{M}_{cks}(Y)$~\cite{Greene:1996cy}. The mirror map is given by~\cite{Candelas:1990rm}
\begin{equation}
	t(z)=\frac{X^1(z)}{X^0(z)}=B+iJ\ ,
\end{equation}
where $X^1(z)$ and $X^0(z)$ are elements of the period vector (see equation~\eqref{eqn:intro_period}), $B=\int_{\mathcal{C}}b$ is the Neveu-Schwarz $B$-field flux around the 2-cycle $\mathcal{C}$ forming a basis of $H_2(Y)$ and $J=\int_{\mathcal{C}}\omega$ is the size of the 2-cycle.

Near $\psi=\infty$ (or $z=0$), we can use equation~\eqref{eqn:sth_pim} and~\eqref{eqn:igamma} to write
\begin{equation}\label{eqn:tlim}
	\begin{split}
	t(z)=&\frac{1}{2\pi i}\frac{\lim_{\epsilon\to 0}\frac{\partial I_{\Gamma}(z,\epsilon)}{\partial \epsilon}}{I_{\Gamma}(z,0)}\\
	=& \frac{1}{2\pi i}\log(z)+h(z)\ ,
	\end{split}
\end{equation}
where
\begin{equation}
	\begin{split}
	h(z)=&\frac{1}{2\pi i}\frac{\sum_{k=0}^{\infty}\left(\lim_{\epsilon\to 0}\frac{\partial g(k,\epsilon)}{\partial\epsilon}\right)z^k}{\sum_{k=0}^{\infty}g(k,0)z^k},\text{ with}\\
	g(k,\epsilon)&={\frac{\Gamma(d_1(k+\epsilon)+1)...\Gamma(d_{n-3}(k+\epsilon)+1)}{\Gamma(k+\epsilon+1)^{n+1}}}\ .		
	\end{split}
\end{equation}
In the limit $z\rightarrow 0$, we get
\begin{equation}\label{eqn:hlimit}
	h(z)\rightarrow \frac{1}{2\pi i} \frac{\gamma\left(n+1-\sum_{i=1}^{n-3} d_i\right)+\mathcal{O}(z)}{1+\mathcal{O}(z)}\stackrel{\eqref{calabicondtn}}{=\joinrel=}\mathcal{O}(z)\ ,
\end{equation}
where $\gamma\sim 0.5772$ is the Euler-Mascheroni constant.

Hence, the mass, $m_{KK}$, mentioned in equation~\eqref{eqn:chap2KK}, in the limit $z\rightarrow 0$, can be calculated using equation~\eqref{eqn:tlim} and equation~\eqref{eqn:hlimit} as
\begin{equation}\label{eqn:massforpsi}
	\begin{split}
	m_{KK}^2\sim\frac{1}{R^2}\sim\frac{1}{\Im(t(z))}&=-\frac{2\pi}{\log(|z|)}+\mathcal{O}\left(\frac{|z|}{\log(|z|)}\right)\\
	&\stackrel{\eqref{eqn:zandpsi}}{=\joinrel=}\frac{2\pi}{\log\left(\mu|\psi|^{n+1}\right)}+\mathcal{O}\left(\frac{1}{|\psi|^{n+1}\log(|\psi|)}\right)\\
	&=\frac{2\pi}{n+1}\frac{1}{\log(|\psi|)}+\mathcal{O}\left(\frac{1}{\log^2(|\psi|)}\right)\ .
	\end{split}
\end{equation}

Next, consider a point $\psi_0>>1$ on the moduli space. For another point $\psi_l>\psi_0$, we can find the distance between $\psi_l$ and $\psi_0$ using the metric in table~\ref{table:glimit},~\ref{table:glimit2} or~\ref{table:glimit3}
\begin{equation}\label{eqn:distforpsi}
	d(\psi_l,\psi_0)\sim\int_{\psi_0}^{\psi_l}\sqrt{\frac{3}{4|\psi|^2\log^2(|\psi|)}}d|\psi|=\frac{\sqrt{3}}{2}\log\left(\frac{\log(|\psi_l|)}{\log(|\psi_0|)}\right)\ .
\end{equation}

Finally, combining equation~\eqref{eqn:distforpsi} and~\eqref{eqn:massforpsi}, we can find the relation between the mass of the KK states and the distance on the moduli space,
\begin{equation}
	m_{KK}^l\sim\frac{1}{\sqrt{\log(|\psi_l|)}}\sim m_{KK}^0 e^{-\frac{1}{\sqrt{3}}d(\psi_l,\psi_0)}\ .
\end{equation}
Of course since the argument of~\cite{Blumenhagen:2018nts} is based on dimensional--  and scale 
arguments relating in particular  the KK scale to the  area as measured by  the K\"ahler parameter  it gives the 
same  exponentially behaviour as the one for the $D2-D0$ brane BPS states, without making statements about their detailed multiplicity.   
In~\cite{Blumenhagen:2018nts}, the authors also construct geodesics on the moduli space to verify the RSDC. 
We will not reproduce the corresponding  calculation for all cases, but qualitatively, one can see that the RSDC
is true for  generic $M$-points. Next, we come to the conceptually more  challenging cases that occur near 
the $\psi=0$ point, which we also call the $s$-point.
\subsection{Light States near the $s$-Point}\label{sec:psi_0}

In this section we try to construct the tower of light states to support the SDC at $s$-point which is at infinite distance for $X_{3,3}$ and $X_{2,2,2,2}$. The candidates are the $3$-branes in type IIB theory which come from reducing the RR 4-form potential $C^{(4)}$ on the $A^I$ cycles. On the $B_I$ cycles, we get their electromagnetic duals. The periods can be used to calculate their central charge, at a point on the moduli space, using
\begin{equation}\label{eqn:centralch}
    Z_\mathbf{q}=e^{K/2}\ \left(\mathbf{q}^{\text{T}}\ \Sigma\ \Pi\right)\ ,
\end{equation}
where $K$ is the K\"{a}hler potential (see equation~\eqref{eqn:Kahler2}), $\mathbf{q}$ is the (electric $+$ magnetic) charge vector, $\Sigma$ is given in equation~\eqref{eqn:sigma} and $\Pi$ is the period vector. 

There is no reason that the 3-branes described above with information of $\mathbf{q}$ are actual, physical BPS states. This is essentially an assumption here. However, what we are able to do here, is checking if these branes remain stable as one approaches the singular point. For that, let's first write down the mass of a physical state. The mass of a BPS 3-brane state is given by~\cite{Ceresole:1995jg}
\begin{equation}
	 m_\mathbf{q}=\left|Z_\mathbf{q}\right|\ .
\end{equation}

The charges $\mathbf{q}$ form a charge lattice and in general, on approaching $\psi=0$, a state will not become massless, but a subset of the charge lattice becomes massless. In the following, we consider $X_{3,3}$ and $X_{2,2,2,2}$, and see what this subset looks like.
\subsubsection{$X_{3,3}$}\label{sec:X33}

Since we are looking at the states when we approach $\psi=0$, we will use the period vector calculated in section~\ref{sec:spoint}. Let's first see the behaviour of the $e^{K/2}$ term in the central charge given in equation~\eqref{eqn:centralch}. Using the period vector, we find (see equation~\eqref{eqn:Kahler2})
\begin{equation}\label{calc:x33K}
    e^{K/2}=-\frac{ 3^{3/4} \pi ^{15/4}}{ \Gamma \left(-\frac{2}{3}\right)^3 \Gamma\left(\frac{1}{6}\right)^{3/2} |\psi|^2 \sqrt{-\log (|\psi|)}}+\mathcal{O}\left(\frac{1}{|\psi|^2(-\log(|\psi|))^{3/2}}\right)
.\end{equation}
Next we want to look at the behaviour of the term $\mathbf{q}^{\text{T}}\ \Sigma\ \Pi$ in equation~\eqref{eqn:centralch}. For a general charge vector $\mathbf{q}$, the structure of periods near the $s$-point gives us a leading (largest) contribution of order $\psi^2\log(\psi)$ which would give a divergent mass when multiplied by $e^{K/2}$ on approaching $\psi=0$. 

Therefore, we next find the subset of the charge lattice which strips the combination $\mathbf{q}^{\text{T}}\ \Sigma\ \Pi$ of logarithmic terms. This gives us
\begin{equation}\label{eqn:calc_qX33}
    \mathbf{q}=\left(
\begin{array}{c}
 x \\
 y \\
 \frac{4 x}{9}+\frac{y}{3} \\
 \frac{x}{3} \\
\end{array}
\right)\ ,
\end{equation}
where $x$ and $y$ are arbitrary integers such that $x\ \text{mod}\ 3=0$ and $(4x+3y)\ \text{mod}\ 9=0$ (quantisation condition). This subset of charge lattice gives us the light states we are looking for and we denote it by
\begin{equation}\label{eqn:Q}
    \mathcal{Q}=\left\{\mathbf{q}\ |\ x,y\in\mathbb{Z},\ x\ \text{mod}\ 3=0,\ (4x+3y)\ \text{mod}\ 9=0\right\}\setminus\left\{(0,0,0,0)^{\text{T}}\right\}
\ .\end{equation}
For the elements in this sub lattice, we find
\begin{equation}\label{eqn:calc_charge33}
    Z_{\mathbf{q}}=e^{K/2}\left(\frac{\sqrt[6]{-1} \sqrt[3]{2} \Gamma \left(-\frac{2}{3}\right)^4 \Gamma \left(\frac{1}{6}\right)
   \left(-14 \sqrt{3} x+3 \left(\sqrt{3}+9 i\right) y\right)}{81 \left(5 \sqrt{3}+3 i\right) \pi ^{7/2}}\psi^2+\mathcal{O}\left(\psi^4\right)\right)    
\ .\end{equation}
This gives us the mass
\begin{equation}\label{eqn:calc_mass33}
    \begin{split}
    m_\mathbf{q}=\left|Z_{\mathbf{q}}\right|&=e^{K/2}\left|\mathbf{q}^{\text{T}}\ \Sigma\ \Pi\right|\\    &=     \frac{-\pi^{1/4} \Gamma \left(-\frac{2}{3}\right) \left| 14 \sqrt{3} x-3 \left(9
   i+\sqrt{3}\right) y\right| }{27\times 2^{2/3}\times 3^{3/4} \sqrt{7\ \Gamma \left(\frac{1}{6}\right)}\sqrt{-\log (|\psi|)}}+\mathcal{O}\left(\frac{|\psi|^2}{\sqrt{-\log(|\psi|)}}\right)\ .
   \end{split}
\end{equation}

Now consider two points $P$ and $Q$ near $\psi=0$ such that $|\psi_P|>|\psi_Q|$, then we can calculate the distance between these two points using metric in table~\ref{table:glimit2}. We get
\begin{equation}
    d_{PQ}\sim\int_{\psi_P}^{\psi_Q}\text{d}|\psi|\sqrt{\frac{1}{4|\psi|^2\log^2(|\psi|)}}=\frac{1}{2}\log\left(\frac{\log(|\psi_Q|)}{\log(|\psi_P|)}\right)
\ .\end{equation}
Combining this with the leading term of mass of the brane calculated in equation~\eqref{eqn:calc_mass33} we can find the ratio of masses at $Q$ and $P$,
\begin{equation}
    \frac{m_{\mathbf{q}}^Q}{m_{\mathbf{q}}^P}\sim\sqrt{\frac{\log(|\psi_P|)}{\log(|\psi_Q|)}}\sim e^{-d_{PQ}}
\ .\end{equation}
Therefore, we observe that for the subset of charge lattice given by $\mathcal{Q}$, the states exponentially become massless as one approaches the $s$-point. If we can show that there are infinitely many such states which are stable, it will be an evidence of the SDC.

In the following, we will construct a subset of the charge lattice $\mathcal{Q}$ which, if assumed to be physically existing close to the singular point, will be stable as one approaches the singular point.

\subsubsection{Stability of Branes in the $X_{3,3}$ Model}\label{sec:stability1}

Let's start with the observation that the central charge of a state can be decomposed as (see equation~\eqref{eqn:calc_qX33})
\begin{equation}\label{eqn:phipsi}
	\begin{split}
		Z_{\mathbf{q}}&=e^{K/2}\left(\mathbf{q}^{\text{T}} \Sigma\Pi\right)\\
					  &=x\Phi+y\Psi\ ,
	\end{split}
\end{equation}
where
\begin{equation}
	\begin{split}
		\Phi&=e^{K/2}\left(\mathbf{q}_1^{\text{T}}\Sigma\Pi\right)\text{ with }\quad \mathbf{q}_1=\left(
\begin{array}{c}
 1 \\
 0 \\
 4/9\\
 1/3 \\
\end{array}
\right)\text{ and}\\
		\Psi&=e^{K/2}\left(\mathbf{q}_2^{\text{T}}\Sigma\Pi\right)\text{ with }\quad \mathbf{q}_2=\left(
\begin{array}{c}
 0 \\
 1 \\
 1/3\\
 0 \\
\end{array}
\right)\ .
	\end{split}
\end{equation}
Note that the states with vanishing mass are linear combination of the vanishing cycles since
\begin{equation}
	\begin{split}
	\frac{\Phi}{e^{K/2}}&=\frac{7 \left(\sqrt{3}+3 i\right) f_1(w) +\left(13 \sqrt{3}-9 i\right)f_2(w) }{12 \left(\sqrt{3}-5 i\right) \pi ^2}\ ,\\
	\frac{\Psi}{e^{K/2}}&=\frac{\left(3+i \sqrt{3}\right) f_1(w)+i \left(\sqrt{3}+3 i\right) f_2(w)}{8 \pi ^2}\ ,
	\end{split}
\end{equation}
where $f_1(w)$ and $f_2(w)$ are given in equation~\eqref{eqn:sX33}.

Now, as noted in~\cite{Seiberg:1994rs}, consider the possible decay process of a BPS 3-brane labelled $A$ with $Z_{\mathbf{q}(A)}=x\Phi+y\Psi$ decaying to states labelled $B_i$ with $Z_{\mathbf{q}(B_i)}=x_i\Phi+y_i\Psi$. Conservation of charge requires that
\begin{equation}
	Z_{\mathbf{q}(A)}=\sum_i Z_{\mathbf{q}(B_i)}\ ,
\end{equation}
however, triangle inequality tells us that
\begin{equation}
	m_{\mathbf{q}(A)}\leq\sum_i m_{\mathbf{q}(B_i)}\ ,
\end{equation}
where $m_{\mathbf{q}}=\left|Z_{\mathbf{q}}\right|$. This inequality is saturated when all the charges $Z_{\mathbf{q}(B_i)}$ are aligned. Hence, for brane $A$ to decay into the branes $B_i$, the charges of the branes $B_i$ must be aligned otherwise the brane $A$ will be stable against this decay. Note that this condition is necessary but not sufficient. Also, this condition defines the walls of marginal stability. 

Assuming that $\Phi$ and $\Psi$ in equation~\eqref{eqn:phipsi} are not aligned, i.e. $\Phi/\Psi$ is not real, the decay of $A$ into $B_i$ (or vice-versa) is only possible if $(x,y)$ is proportional to $(x_i,y_i)$. In other words, if $(x,y)=(nk,nl)$ for some $n\in\mathbb{Z}$ and $k,l\in\mathbb{Z}$ satisfying $k\text{ mod }3=0$ and $(4k+3l)\text{ mod }9=0$ (quantisation condition). On the contrary, if $x$ and $y$ are relative primes, the state will be stable against such a decay. Since there are infinitely many relative primes in the set $\mathcal{Q}$ given in equation~\eqref{eqn:Q}, we have infinitely many light stable states if $\Phi/\Psi\notin\mathbb{R}$.

\begin{figure}[t]
    \centering
\begin{overpic}[scale=0.4]{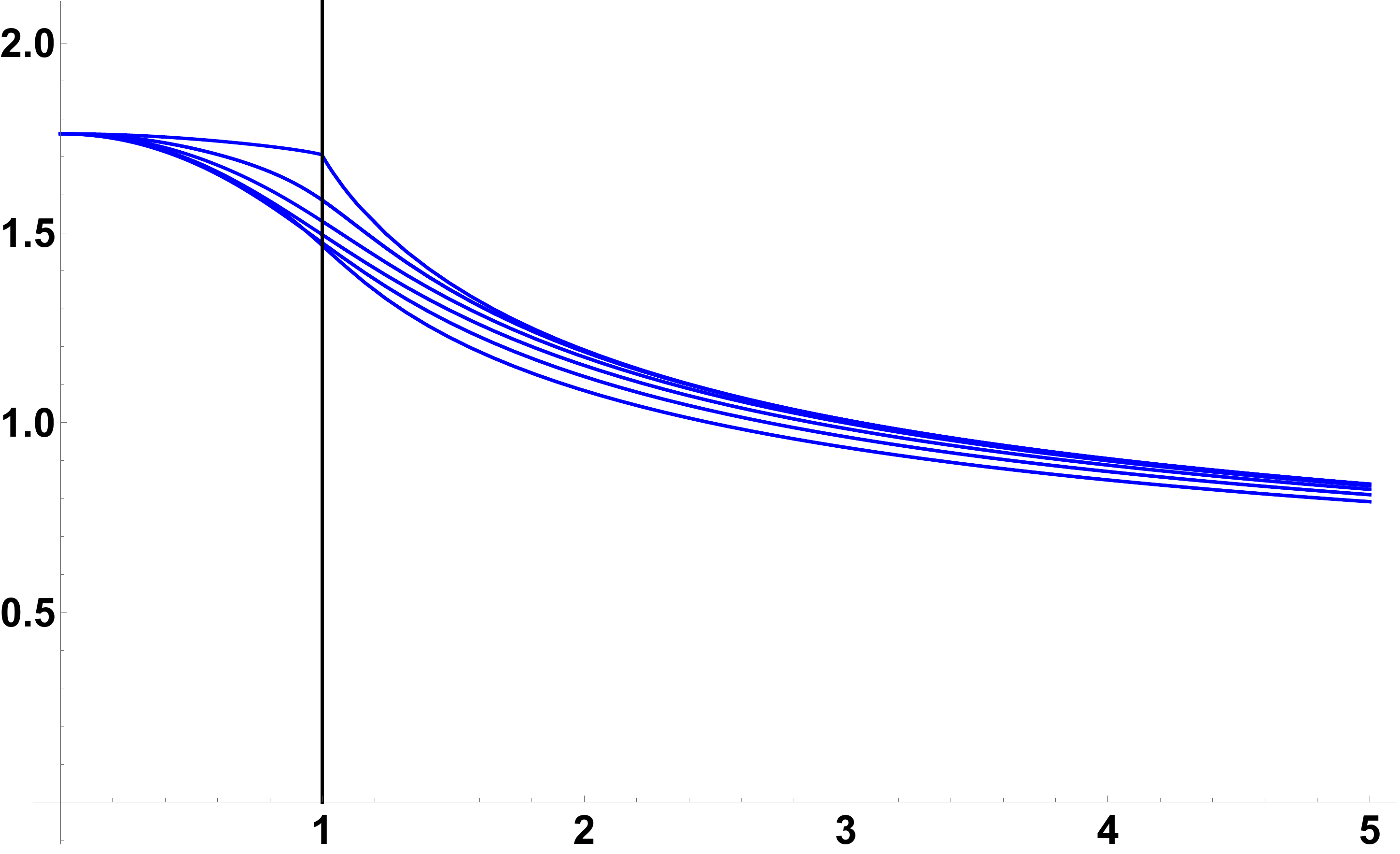}
    \put (50,-1) {$|\psi|$}
    \put (-13,36) {$\text{arg}(\Phi/\Psi)$}
\end{overpic}
 \caption{Plot of $\text{arg}(\Phi/\Psi)$ against $|\psi|$ for $\text{arg}(\psi)=2\pi k/36,\ k=\{0,1,2,3,4,5,6\}$. The blue line with cusp at $|\psi|=1$ is for $k=\{0,6\}$ and rest of the blue lines are for $k=1,2,3,4,5$ in increasing order of height.}
 \label{figm:ratiopsiphi}
\end{figure}

With the knowledge of periods on the whole moduli space, we can indeed compute $\Phi$ and $\Psi$. More importantly, we look at the difference in the argument of $\Phi$ and $\Psi$ or $\text{arg}(\Phi/\Psi)$. The plot is given in figure~\ref{figm:ratiopsiphi}. In the figure, we plot $\text{arg}(\Phi/\Psi)$ against $|\psi|$ for $\text{arg}(\psi)=2\pi k/36,\ k=\{0,1,2,3,4,5,6\}$. The blue line with cusp at $|\psi|=1$ is for $k=\{0,6\}$ (conifold points) and rest of the blue lines are for $k=1,2,3,4,5$ in increasing order of height. As $|\psi|\rightarrow\infty$, $\text{arg}(\Phi/\Psi)\rightarrow 0$. Clearly, at no point on the moduli space, other than $\psi=\infty$, $\text{arg}(\Phi/\Psi)=0$ or $\pi$. Or, at no point $\Phi$ and $\Psi$ align, hence, the states with relatively prime $(x,y)$ mentioned above will be stable.

\begin{figure}[t]
    \centering
\begin{overpic}[scale=0.4]{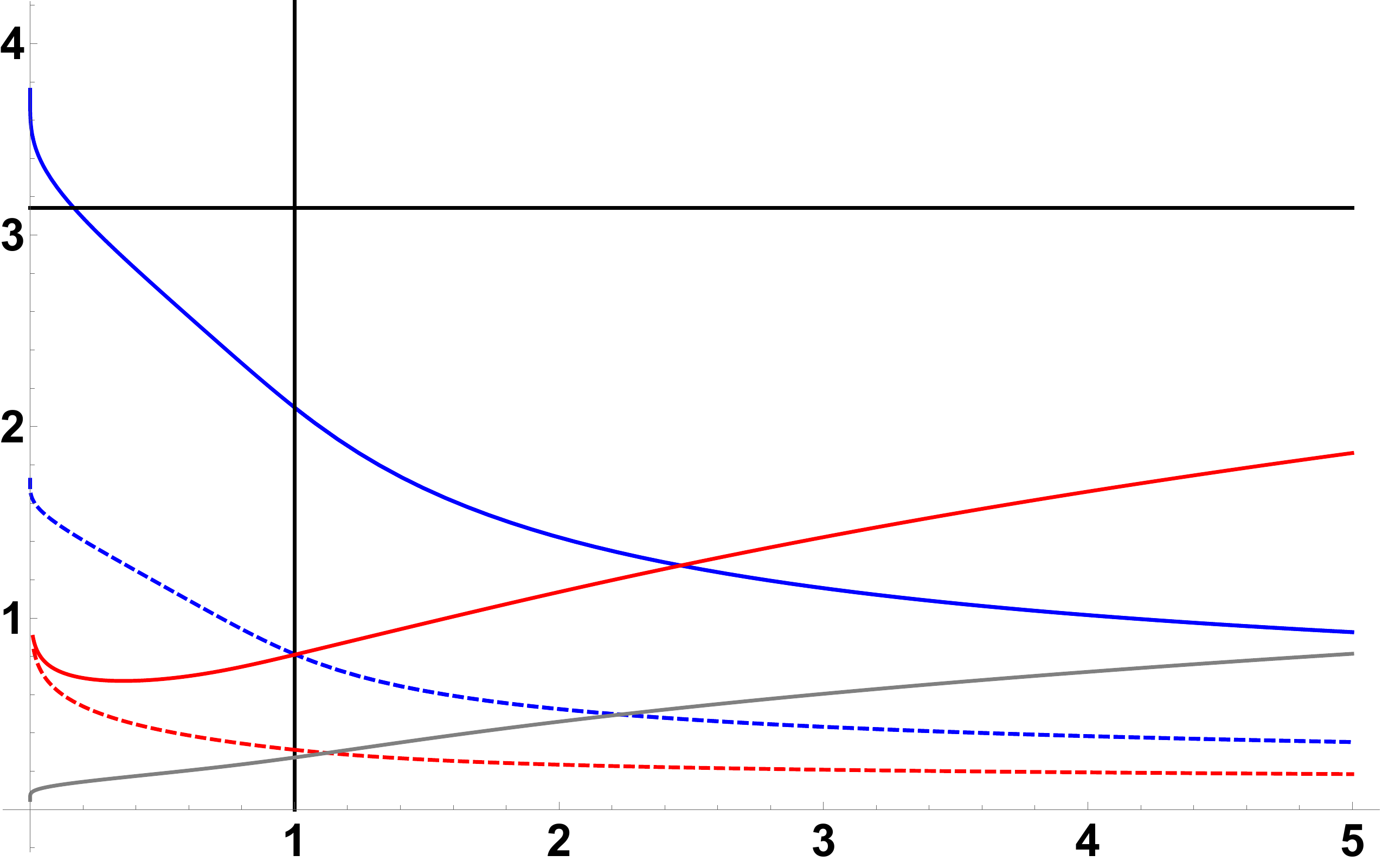}
    \put (50,-1) {$|\psi|$}
\end{overpic}
 \caption{Mass and $\text{arg}(\Phi/\Xi)$ against $|\psi|$ for $\text{arg}(\psi)=2\pi/18$. Blue lines show $\text{arg}(\Phi/\Xi)$, red lines show the mass. Dashed is for $\tilde{\mathbf{q}}_1=(0,1,0,0)^{\text{T}}$ and thick is for $\tilde{\mathbf{q}}_2=(1,1,1,0)^{\text{T}}$. Grey line shows $|\Phi|$.}
 \label{fig:ratiomassive}
\end{figure}

It is important to point out that above we have only considered light states and shown that there are infinitely many light states which are stable against decay into each other. The states which are not of the form that belongs to $\mathcal{Q}$ have divergent mass as we approach $\psi=0$, let's call them massive states. The non-triviality enters when we move away from $\psi=0$ which leads to the light states gaining mass and the massive states becoming light. It turns out that performing calculations similar to above, we can see that the light states can decay into massive ones and vice-versa away from $\psi=0$. 

Let's start with defining 
\begin{equation}
	\Xi=e^{K/2}\left(\tilde{\mathbf{q}}^{\text{T}}\Sigma\Pi\right)
\end{equation} 
where $\tilde{\mathbf{q}}\notin\mathcal{Q}$. In figure~\ref{fig:ratiomassive}, the plot shows the $\text{arg}(\Phi/\Xi)$ for $\tilde{\mathbf{q}}_1=(0,1,0,0)^{\text{T}}$ in dashed blue line and for $\tilde{\mathbf{q}}_2=(1,1,1,0)^{\text{T}}$ in thick blue line. Also, in dashed red and thick red, we have plotted $m=\left|\Xi\right|$ for $\tilde{\mathbf{q}}_1$ and $\tilde{\mathbf{q}}_2$ respectively and grey line shows $\left|\Phi\right|$\footnote{$\Phi$ is charge for $\mathbf{q}_1=(1,0,4/9,1/3)^{\text{T}}$, hence $|\Phi|$ represents $1/9^{\text{th}}$ of the mass of a state given by $\mathbf{q}=(9,0,4,3)^{\text{T}}$ (quantisation condition).}. Note that for $\tilde{\mathbf{q}}_1$, the blue line crosses the black line ($\text{arg}(\Phi/\Xi)=\pi$) near $\psi=0$ aligning the two charges whereas for $\tilde{\mathbf{q}}_2$ it never happens, also, at this point, the light state and heavy state have comparable mass. Hence, we can conclude that there are lines of marginal stability near $\psi=0$ if all the states are taken into account.

This is however not a problem for us because, in the limit of approaching $\psi=0$, the heavy states become infinitely heavy and light states become massless not allowing any decay to take place between the two classes and the light states remain stable. And hence, the light states with relatively prime $x$ and $y$, under the assumption of stability close to $\psi=0$ advocate for SDC. Also, we have shown that for a state which becomes massive at singularity, there is always a finite radius around the $K$-point in which an infinite number of light states are stable against decay into each other as well as the above mentioned charge.

Next, we will do a similar analysis for $X_{2,2,2,2}$.
\subsubsection{$X_{2,2,2,2}$}\label{sec:X2222}

It turns out that for $X_{2,2,2,2}$, everything follows just like it did for $X_{3,3}$. Again, using the period vector calculated in section~\ref{sec:spoint} we can find $e^{K/2}$,

\begin{equation} e^{K/2}=\frac{\sqrt{3}\pi^{3/2}}{64|\psi|^4(-\log(|\psi|))^{3/2}}+\frac{3\sqrt{3}\pi^{3/2}\log(2)}{128|\psi|^4(-\log(|\psi|))^{5/2}}+\mathcal{O}\left(\frac{1}{|\psi|^4(-\log(|\psi|))^{7/2}}\right)\ 
.\end{equation}

Next, we look at the behaviour of $\mathbf{q}^{\text{T}}\ \Sigma\ \Pi$ term in equation~\eqref{eqn:centralch}. For general charge $\mathbf{q}$, now the leading (largest) contribution is of order $\psi^4\log^3(\psi)$ with sub-leading contributions of order $\psi^4\log^2(\psi)$ and $\psi^4\log(\psi)$ respectively. In contrast to the case in the last section, we just need to eliminate terms with logarithmic order two or higher to get a convergent, vanishing mass (central charge). We find that this is accomplished using the subset of charge lattice consisting of charges of the form
\begin{equation}\label{eqn:calc_q2222}
    \mathbf{q}=\left(
\begin{array}{c}
 x \\
 y \\
 \frac{x}{2}+\frac{y}{4} \\
 \frac{x}{4} \\
\end{array}
\right)\ ,
\end{equation}
where $x$ and $y$ are arbitrary integers such that $x\ \text{mod}\ 4=0$ and $y\ \text{mod}\ 4=0$ (quantisation condition). We denote this subset by
\begin{equation}\label{eqn:calc_2222subQ}
    \mathcal{Q}=\{\mathbf{q}\ |\ x,y\in\mathbb{Z},\ x\ \text{mod}\ 4=0,\ (2x+y)\ \text{mod}\ 4=0\}\setminus \left\{(0,0,0,0)^{\text{T}}\right\}
\ .\end{equation}
For these charges, we find
\begin{equation}\label{eqn:calc_charge2222}
    Z_{\mathbf{q}}=e^{K/2}\left(-\frac{4 i y }{\pi }\psi^4\log (\psi )+\frac{2 i y \log (4)-x}{\pi }\psi ^4+\mathcal{O}\left(\psi^{12}\right)\right)
\ .\end{equation}

Thus, the mass of a BPS state is
\begin{equation}
    m_{\mathbf{q}}=e^{K/2} |\mathbf{q}^{\text{T}}\ \Sigma\ \Pi|= \frac{\sqrt{3\pi}\ |y|}{16\sqrt{-\log(|\psi|)}}+\mathcal{O}\left(\frac{1}{(-\log(|\psi|))^{3/2}}\right)
\ .\end{equation}
Note that the equation above has no $x$ dependence. This doesn't mean that the lattice is one dimensional, it merely means that the $x$ dependence is sub-leading which can be seen from equation~\eqref{eqn:calc_charge2222}. This would mean that the spacing of mass states is different in the $x$ and $y$ direction. That is, for fixed $x$ and different $y$, we will have larger mass difference than that of different $x$ and fixed $y$.

Therefore, just like in the last section, using the metric from table~\ref{table:glimit3}, we find that the ratio of masses at $Q$ and $P$ for $1\gg|\psi_P|>|\psi_Q|$ is
\begin{equation} \frac{m_{\mathbf{q}}^Q}{m_{\mathbf{q}}^P}\sim\sqrt{\frac{\log(|\psi_P|)}{\log(|\psi_Q|)}}\sim e^{-\frac{1}{\sqrt{3}}d_{PQ}}\ ,
\end{equation}
as we approach the singularity at $\psi=0$ following the behaviour predicted by SDC. 

\subsubsection{Stability of Branes in the $X_{2,2,2,2}$ Model}\label{sec:stability2}

The stability arguments follow analogous to the ones in the last section. We first decompose the central charge as
\begin{equation}\label{eqn:phipsi2}
	\begin{split}
		Z_{\mathbf{q}}&=e^{K/2}\left(\mathbf{q}^{\text{T}}\Sigma\Pi\right)\\
					  &=x\Phi+y\Psi
	\end{split}
\end{equation}
where
\begin{equation}
	\begin{split}
		\Phi&=e^{K/2}\left(\mathbf{q}_1^{\text{T}}\Sigma\Pi\right)\text{ with }\quad \mathbf{q}_1=\left(
\begin{array}{c}
 1 \\
 0 \\
 1/2\\
 1/4 \\
\end{array}
\right)\text{ and}\\
		\Psi&=e^{K/2}\left(\mathbf{q}_2^{\text{T}}\Sigma\Pi\right)\text{ with }\quad \mathbf{q}_2=\left(
\begin{array}{c}
 0 \\
 1 \\
 1/4\\
 0 \\
\end{array}
\right)\ .
	\end{split}
\end{equation}

\begin{figure}[t]
    \centering
\begin{overpic}[scale=0.4]{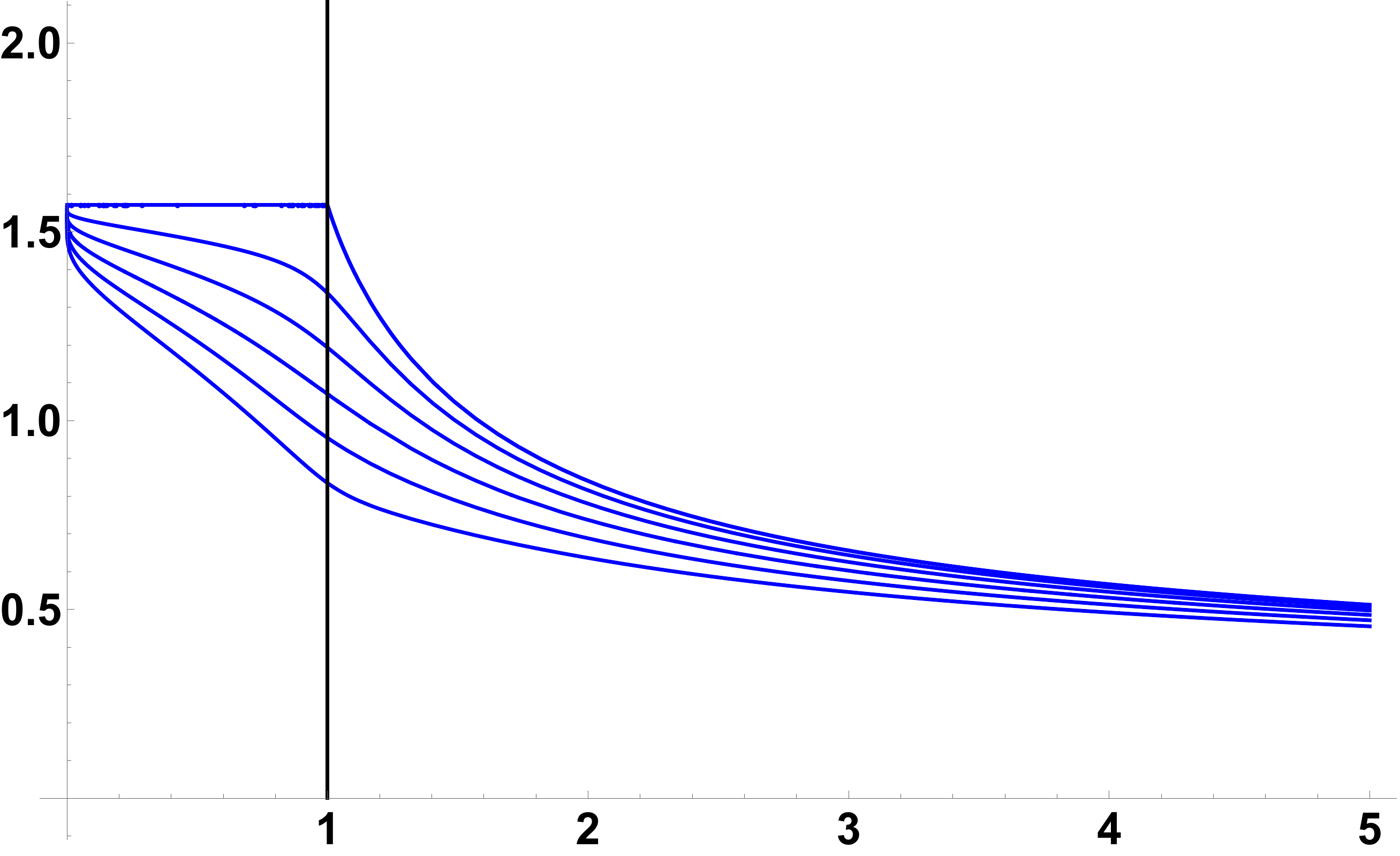}
    \put (50,-1) {$|\psi|$}
    \put (-13,36) {$\text{arg}(\Phi/\Psi)$}
\end{overpic}
 \caption{Plot of $\text{arg}(\Phi/\Psi)$ against $|\psi|$ for $\text{arg}(\psi)=2\pi k/48,\ k=\{0,1,2,3,4,5,6\}$. The blue line with cusp at $|\psi|=1$ is for $k=\{0,6\}$ and rest of the blue lines are for $k=1,2,3,4,5$ in increasing order of height.}
 \label{fig:ratio2222}
\end{figure}

In figure~\ref{fig:ratio2222}, we plot $\text{arg}(\Phi/\Psi)$ against $|\psi|$ for $\text{arg}(\psi)=2\pi k/48$, $k=\{0,1,2,3,4,5,6\}$. Again, at not point on the moduli space, except $\psi=\infty$, $\Phi$ and $\Psi$ align. Hence, for relatively prime $x$ and $y$, the light states are stable everywhere under decaying into each other.

\begin{figure}[t]
    \centering
\begin{overpic}[scale=0.4]{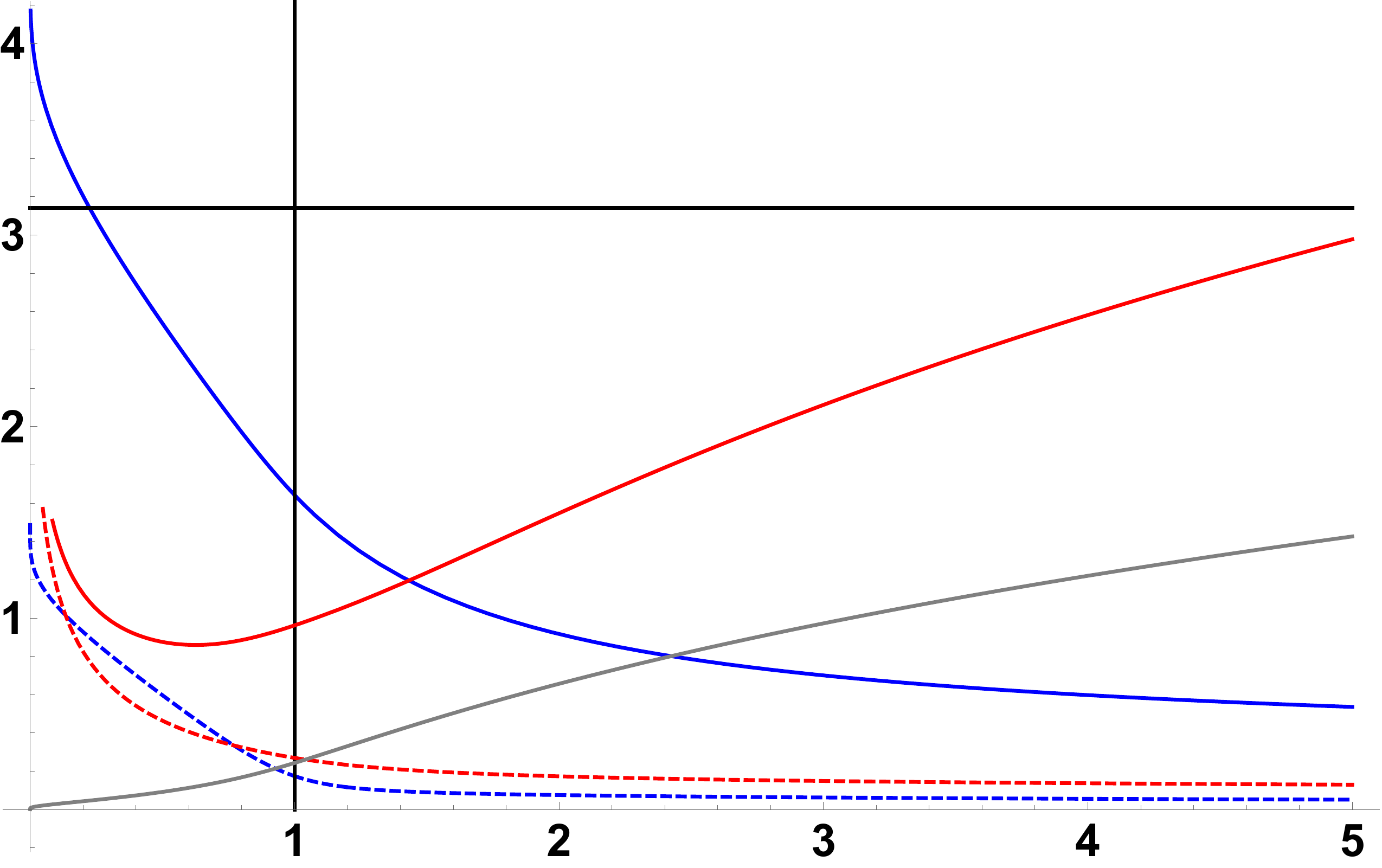}
    \put (50,-1) {$|\psi|$}
\end{overpic}
 \caption{Mass and $\text{arg}(\Phi/\Xi)$ against $|\psi|$ for $\text{arg}(\psi)=2\pi/24$. Blue lines show $\text{arg}(\Phi/\Xi)$, red lines show the mass. Dashed is for $\tilde{\mathbf{q}}_1=(0,1,0,0)^{\text{T}}$ and thick is for $\tilde{\mathbf{q}}_2=(1,1,1,0)^{\text{T}}$. Grey line shows $|\Phi|$.}
 \label{fig:ratiomass2222}
\end{figure}

Next, we define
\begin{equation}
	\Xi=e^{K/2}\left(\tilde{\mathbf{q}}^{\text{T}}\Sigma\Pi\right)\ ,
\end{equation}
where $\tilde{\mathbf{q}}\notin\mathcal{Q}$. In figure~\ref{fig:ratiomass2222}, we have $\text{arg}(\Phi/\Xi)$ for $\tilde{\mathbf{q}}_1=(0,1,0,0)^{\text{T}}$ in dashed blue line and for $\tilde{\mathbf{q}}_2=(1,1,1,0)^{\text{T}}$ in thick blue line. Respective red lines show the $m=|\Xi|$ and grey line shows $|\Phi|$. Note again, there exists charges with line of marginal stability which can decay light states into massive states (or vice versa).

However, just like before, this is not a problem assuming that there are stable light states near $\psi=0$ with relatively 
prime $x$ and $y$.

There are two comment in order. Unlike light states for $X_{3,3}$ which have a  
finite order monodromy around  the $K$-point, the light states of the $X_{2,2,2,2}$ model 
have infinite order monodromy so the argument of~\cite{Grimm:2018,Grimm:2018cpv} could apply. 
But there is a much better argument confirming the stability of infinite states including their degenerations. 
Using the results of~\cite{Huang:2006hq} one  can calculate the $\mathcal{F}_g$ at the new $M$-points. 
As it turns out, even though the leading order of the periods is different, one can make a K\"ahler transformation 
so that the genus one prepotential $\mathcal{F}_0$ in the flat coordinates $s=Y_1/Y_0$ is 
given by the ratio of the single logarithmic by the analytic period becomes exactly the same as the one at the 
original $M$-point. To see this, we note that the period calculated in section~\ref{sec:spoint} near $s$-point for $X_{2,2,2,2}$ can be written as
\begin{equation}
 \Pi_s(w)=4\sqrt{\tilde{w}}\left(
\begin{array}{cccc}
 -4 & 0 & 0 & 0 \\
 0 & 4 & 0 & 0 \\
 -2 & 1 & -4 & 0 \\
 -1 & 0 & 0 & 4 \\
\end{array}
\right)\Pi_M(\tilde{w})\ ,
\end{equation}
where $\tilde{w}=2^{16}w$ and $\Pi_M$ is from section~\ref{sec:mpoint}. The same is true for the higher genus potentials. Then we can perform the Schwinger loop calculation and determine the BPS indices at the second $M$-points in the same way then at the original $M$-point only that the charges of the light states are not given by the $D2$ and $D0$ brane charges at the original $M$-points. However we can fix the boundary conditions for the ${\cal F}_g$ so that  the BPS indices  of the  
corresponding bound states are exactly the same as the ones at the original  $M$-point calculated in \cite{Huang:2006hq} and displayed in Table 5.             
\begin{table}
\begin{centering}
\begin{tabular}{|r|rrrrrrr|}
\hline
g &$\beta=1$ &$\beta=2 $&$\beta=3 $&$\beta=4 $&$\beta=5 $&$\beta=6$&$\beta=7$\\
\hline
0&  512& 9728& 416256& 25703936& 1957983744& 170535923200& 16300354777600   \\
1& 0& 0& 0& 14752& 8782848& 2672004608& 615920502784 \\
2& 0& 0& 0& 0& 0& 1427968& 2440504320\\
3& 0&0& 0& 0& 0& 0& 86016 \\
4&0& 0& 0& 0& 0& 0& 0  \\
\hline
\end{tabular}
\caption{ $I_g^\beta$ for the degree (2,2,2,2) complete intersection in $\mathbb{P}^7$ at the second $M$-point ($\psi=0$).}
\end{centering}
\end{table}
\subsection{Other One Parameter Hypergeometric Systems}\label{sec:othermodels}
With the help of Meijer G-functions and knowledge of periods around the  $M$-point, we find the structure of the metric on approaching the  $s$-point for all the $14$ hypergeometric one parameter cases and also find the lattice subspace for cases with infinite distance. We present the results in table~\ref{tab:allmetric2} where we also mention the constants $\mu$, $\kappa$ and $c_2\cdot D$ needed to find the periods around  $M$-point (see equation~\eqref{eqn:igamma}). We show the structure of the metric when we approach  $s$-point and also the distance to the  $s$-point from a non-singular point on the moduli space.

Further, for all examples where the distance to $s$-point in infinite, we can find the subset of charge lattice leading to massless states on approaching the  $s$-point. The general form of charge vector for each case is given below.
\begin{equation}
	\begin{split}
		 & X_{2,2,2,2}(1^8): \mathbf{q}=\left(
		\begin{array}{c}
		 x \\
		 y \\
		 \frac{x}{2}+\frac{y}{4} \\
		 \frac{x}{4} \\
		\end{array}
		\right),\text{    }\qquad\quad
		X_{3,3}(1^6): \mathbf{q}=\left(
		\begin{array}{c}
		 x \\
		 y \\
		 \frac{4x}{9}+\frac{y}{3} \\
		 \frac{x}{3} \\
		\end{array}
		\right),\\ \null\\
		 &X_{4,4}(1^42^2): \mathbf{q}=\left(
		\begin{array}{c}
		 x \\
		 y \\
		 \frac{x}{2}+\frac{y}{2} \\
		 \frac{x}{2} \\
		\end{array}
		\right)\ \text{and}\qquad
		 X_{6,6}(1^22^23^2):\mathbf{q}=\left(
		\begin{array}{c}
		 x \\
		 y \\
		 y \\
		 x \\
		\end{array}
		\right)\ .
	\end{split}
\end{equation}

\acknowledgments
We would like to thank Hans Jockers for useful discussions. We would also like to thank Andreas Gerhardus for proofreading the paper. A.J. would like to thank the Bonn-Cologne Graduate School of Physics and Astronomy (BCGS) for their financial support.

\appendix

\section{Meijer G-Functions}\label{app:mgfunction}

Here, we briefly describe how one goes about solving the following differential equation
\begin{equation}\label{eqn:app_zinforig}
    \left(\vartheta^4-\mu\frac{1}{w}\prod_{k=1}^{4}(\vartheta-a_k)\right)f(w)=0\ ,
\end{equation}
where $\vartheta=w\frac{d}{dw}$. This equation can be rewritten as
\begin{equation}\label{eqn:app_zinf}
    \left(\prod_{k=1}^{4}(\theta-a_k)-x\theta^4\right)f(x)=0\ ,
\end{equation}
where $x=\frac{w}{\mu}$ and $\theta=x\frac{d}{dx}$. The differential equation~\eqref{eqn:app_zinf} is an example of a general differential equation of type
\begin{equation}\label{eqn:app_meidiff}
    \left(\prod_{k=1}^q (\theta-b_k)-x\prod_{k=1}^p(\theta+1-c_k)\right)f(x)=0\ ,
\end{equation}
whose family of solutions was introduced by Meijer~\cite{zbMATH02526339} and are denoted by
\begin{equation}\label{eqn:app_meisoln}
    f(x)=G_{p,q}^{m,n}\left((-1)^{m+n+p}x|
    \begin{array}{c}
     c_1',...,c_p' \\
     b_1',...,b_q' \\
    \end{array}
    \right),\quad\quad 0< m\leq q,\ 0< n\leq p\ ,
\end{equation} 
where $\{c_k'\}$ and $\{b_k'\}$ are permutations of the original $\{c_k\}$ and $\{b_k\}$.~\cite{beals2013meijer} gives a nice introduction to these functions which are known as \textit{Meijer G-functions}. For our case, on comparing equation~\eqref{eqn:app_zinf} and~\eqref{eqn:app_meidiff}, we find
\begin{equation}
    \begin{array}{c}
        p=q=4\\
        c_k=1\\
        b_k=a_k\ .
    \end{array}
\end{equation}
On counting the total number of solutions, we find that there are $4\times 4$ solutions due to choices of  $m$ and $n$, and for each of them, we will have multiple solutions due to the permutations of $\{c_k\}$ and $\{b_k\}$ (or one if all $b_k$ are equal).

Note that the differential equation we have is of the fourth order, which means that there should only be four independent solutions. This is consistent with the larger number of solutions above because most of the solutions constructed using Meijer functions are linearly dependent. To see this more clearly, we look at the Mellin-Barnes type integral representation of these functions. We define~\cite{bateman1953higher}
 \begin{equation}\label{eqn:app_barnes}
    G_{p,q}^{m,n}\left(x|
    \begin{array}{c}
     c_1,...,c_p \\
     b_1,...,b_q \\
    \end{array}\right)=\frac{1}{2\pi i}\int_{\mathcal{C}}\frac{\prod_{k=1}^m\Gamma(b_k-s)\prod_{k=1}^n\Gamma(1-c_k+s)}{\prod_{k=m+1}^q\Gamma(1-b_k+s)\prod_{k=n+1}^p\Gamma(c_k-s)}x^s ds
 \ .\end{equation}
Here, empty product is taken to be $1$. The contour $\mathcal{C}$ is shown in figure~\ref{fig:app_contour}. Here, with blue, we have marked the poles of $\Gamma(b_k-s)$, $k=1,...,m$ and with green, we have poles of $\Gamma(1-c_k+s)$, $k=1,...,n$. There are two ways we can close this contour:
\begin{enumerate}
    \item $\mathcal{C}$ is closed to the right encircling all of the poles of $\Gamma(b_k-s)$, $k=1,...,m$, once in the negative direction and none of the poles of $\Gamma(1-c_k+s)$, $k=1,...,n$. This will converge for $|x|<1$.
    \item $\mathcal{C}$ is closed to the left encircling all of the poles of $\Gamma(1-c_k+s)$, $k=1,...,n$, once in the positive direction and none of the poles of $\Gamma(b_k-s)$, $k=1,...,n$. This will converge for $|x|>1$.
	\end{enumerate}
    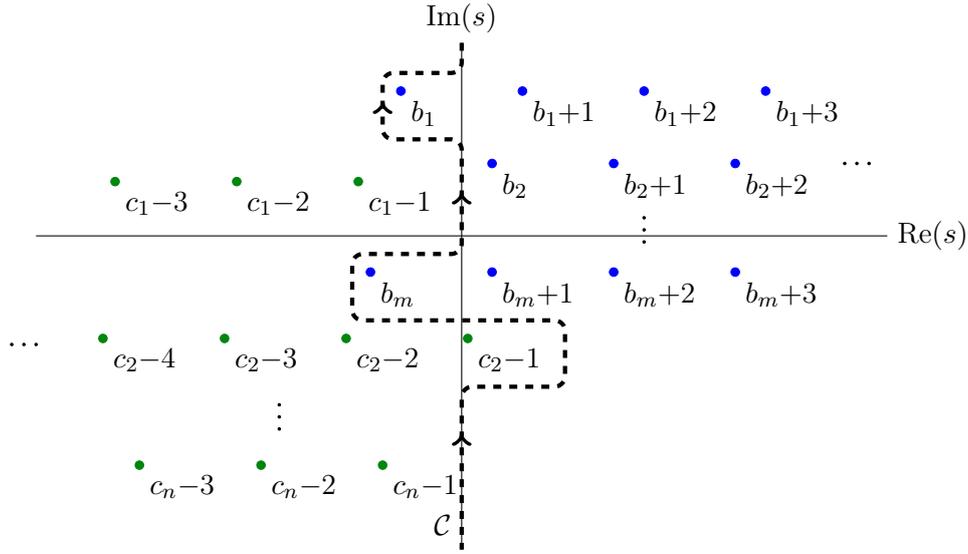
\begin{figure}
        \centering
    \begin{tikzpicture}[scale=0.8]
        \draw[-] (0,-4.4) -- (0,4);
        \node [above] at (0,4) {Im$(s)$};
        \draw[-] (-7,0.8) -- (7,0.8);
        \node [right] at (7,0.8) {Re$(s)$}; 

        \foreach \x in {0,...,3}
            {\draw[blue,fill] (-1+2*\x,3.2) circle [radius=0.07];
            \node [below right] at (-1+2*\x,3.2) {\ifthenelse{\x=0}{$b_1$}{$b_1+$\x}};}
        \foreach \x in {0,...,2}
            {\draw[blue,fill] (0.5+2*\x,2) circle [radius=0.07];
            \node [below right] at (0.5+2*\x,2) {\ifthenelse{\x=0}{$b_2$}{$b_2+$\x}};}
        \foreach \x in {0,...,3}
            {\draw[blue,fill] (-1.5+2*\x,-0.9+1.1) circle [radius=0.07];
            \node [below right] at (-1.5+2*\x,-0.9+1.1) {\ifthenelse{\x=0}{$b_m$}{$b_m+$\x}};}
        \foreach \x in {1,...,3}
            {\draw[oli,fill] (-0.7-2*\x+1,1+0.7) circle [radius=0.07];
            \node [below right] at (-0.7-2*\x+1,1+0.7) {$c_1-$\x};}
        \foreach \x in {1,...,4}
            {\draw[oli,fill] (+1.1-2*\x+1,-2+1.1) circle [radius=0.07];
            \node [below right] at (+1.1-2*\x+1,-2+1.1) {$c_2-$\x};}
        \foreach \x in {1,...,3}
            {\draw[oli,fill] (+0.1-2*\x+0.6,-3) circle [radius=0.07];
            \node [below right] at (+0.1-2*\x+0.6,-3) {$c_n-$\x};}
            
        \foreach \x in {0,...,2}
            {\draw[black,fill] (+6.3+0.2*\x,2) circle [radius=0.02];}
        \foreach \x in {0,...,2}
            {\draw[black,fill] (-6-0.2*\x-1,-1) circle [radius=0.02];}

        \foreach \x in {0,...,2}
            {\draw[black,fill] (3,1.1-0.2*\x) circle [radius=0.02];}
        \foreach \x in {0,...,2}
            {\draw[black,fill] (-3,-2-0.2*\x) circle [radius=0.02];}
        
        \draw[dashed,rounded corners,ultra thick] (0,4)--(0,3.5)--(-1.3,3.5)--(-1.3,2.4)--(0,2.4)--(0,-0.5+1)--(-1.8,-0.5+1)--(-1.8,-1.6+1)--(1.7,-1.6+1)--(1.7,-2.6+1-0.1)--(0,-2.6+1-0.1)--(0,-3.6)--(0,-4.4);
        \node[left] at (0,-4) {$\mathcal{C}$};
        
        \draw [->,ultra thick] (0,1.5) -- (0,1.501);
        \draw [->,ultra thick] (0,-2.5) -- (0,-2.5+0.001);
        \draw [->,ultra thick] (-1.3,3) -- (-1.3,3.001);

    \end{tikzpicture}
\caption{The contour $\mathcal{C}$ with the poles of $\Gamma(b_k-s)$, $k=1,...,m$ marked with blue and the poles of $\Gamma(1-c_k+s)$, $k=1,...,n$ marked with green.}
\label{fig:app_contour}
	\end{figure}

The following are some useful observations. The family of solutions we had before, equation~\eqref{eqn:app_meisoln}, will have redundancies since we can commute Gamma functions in the numerator or in the denominator of right side of the equation~\eqref{eqn:app_barnes}. The integral representation is also useful in analytic continuation of the function from a region where $|x|<1$ to a region where $|x|>1$ or vice versa by closing the contour on different sides.

For $p=q$, if no two $b_j$, $j=1,...,m$ differ by an integer, then all poles of~\eqref{eqn:app_barnes} will be of first order if we close the contour to the right, i.e. for $|x|<1$, which allows us to write the Meijer functions in terms of hypergeometric functions $\,_aF_b$~\cite{bateman1953higher}
\begin{equation}
    \begin{split}
    G_{p,q}^{m,n}\left(x|
    \begin{array}{c}
     c_1,...,c_p \\
     b_1,...,b_q \\
    \end{array}\right)=&\sum_{h=1}^m\frac{{\prod\limits_{j=1}^m} \!' \Gamma\left(b_j-b_h\right)\prod\limits_{j=1}^{n}\Gamma\left(1+b_h-c_j\right)}{\prod\limits_{j=m+1}^q\Gamma\left(1+b_h-b_j\right)\prod\limits_{j=n+1}^{p}\Gamma\left(c_j-b_h\right)}x^{b_h}\\
    & \times\,_pF_{q-1}(1+b_h-c_1,...,1+b_h-c_p;\\
    & \qquad\quad\quad1+b_h-b_1,...,*,...,1+b_h-b_q;\left(-1\right)^{p-m-n}x)\ ,
\end{split}
\end{equation}
where prime in $\prod '$ indicates that we omit term $\Gamma\left(b_h-b_h\right)$ and the asterisk indicates omission of $1+b_h-b_h$ term. The hypergeometric functions can be written as a power series
\begin{equation}
    \,_uF_v(a_1,...,a_u;b_1,...,b_v;x)=\sum_{n=0}^{\infty}\frac{(a_1)_n...(a_u)_n}{(b_1)_n...(b_v)_n}\frac{x^n}{n!}\ ,
\end{equation}
where
\begin{equation}
    (a)_k=\frac{\Gamma\left(a+k\right)}{\Gamma\left(a\right)},\quad (a)_0=1
\ .\end{equation}

Let us now use the method stated above to explicitly find the solutions of equation~\eqref{eqn:app_zinforig} for the cases discussed in this work. Following are the results ($x=\frac{w}{\mu}$): 

    1. For $X_{4,2}(1^6)$, we have $(a_1,a_2,a_3,a_4)=(1/4,1/2,1/2,3/4)$ and $\mu=2^{10}$.
    
    Family of solutions:
    \begin{equation}
        G_{4,4}^{m,n}\left((-1)^{m+n+4}\frac{w}{2^{10}}|
            \begin{array}{c}
             1,1,1,1 \\
             \frac{1}{4}',\frac{1}{2}',\frac{1}{2}',\frac{3}{4}' \\
            \end{array}
            \right)
    \ .\end{equation}
    Four linearly independent solutions:
    \begin{equation}\label{app:eqn_X42}
        \begin{array}{c}
        f_1(w)=G_{4,4}^{1,4}\left(-\frac{w}{2^{10}}|
        \begin{array}{c}
         1,1,1,1 \\
         \frac{1}{4},\frac{1}{2},\frac{1}{2},\frac{3}{4} \\
        \end{array}
        \right)\xrightarrow[]{|w|<2^{10}}\frac{ \Gamma \left(\frac{1}{4}\right)^4 }{ \sqrt{\pi} \Gamma
           \left(\frac{3}{4}\right)^2}\left(\frac{-w}{2^{10}}\right)^{1/4} \,
           _4F_3\left(\frac{1}{4},\frac{1}{4},\frac{1}{4},\frac{1}{4};\frac{1}{2},\frac{3}{4},\frac{3}{4};\frac{w}{2^{10}}\right)\ ,\\
        \null\\
        f_2(w)=G_{4,4}^{1,4}\left(-\frac{w}{2^{10}}|
        \begin{array}{c}
         1,1,1,1 \\
         \frac{1}{2},\frac{1}{4},\frac{1}{2},\frac{3}{4} \\
        \end{array}
        \right)\xrightarrow[]{|w|<2^{10}}\frac{\pi^2  }{ \Gamma \left(\frac{3}{4}\right)
           \Gamma \left(\frac{5}{4}\right)}\left(\frac{-w}{2^{10}}\right)^{1/2} \, _4F_3\left(\frac{1}{2},\frac{1}{2},\frac{1}{2},\frac{1}{2};\frac{3}{4},1,\frac{5}{4};\frac{w}{2^{10}}\right)\ ,\\
           \null\\
        f_3(w)=G_{4,4}^{1,4}\left(-\frac{w}{2^{10}}|
        \begin{array}{c}
         1,1,1,1 \\
         \frac{3}{4},\frac{1}{4},\frac{1}{2},\frac{1}{2} \\
        \end{array}
        \right)\xrightarrow[]{|w|<2^{10}}\frac{ \Gamma \left(\frac{3}{4}\right)^4 }{  \Gamma\left(\frac{5}{4}\right)^2\Gamma\left(\frac{3}{2}\right)}\left(\frac{-w}{2^{10}}\right)^{3/4} \,   _4F_3\left(\frac{3}{4},\frac{3}{4},\frac{3}{4},\frac{3}{4};\frac{5}{4},\frac{5}{4},\frac{3}{2};\frac{w}{2^{10}}\right)\ ,\\
           \null\\
        f_4(w)=G_{4,4}^{2,4}\left(\frac{w}{2^{10}}|
        \begin{array}{c}
        1,1,1,1 \\
        \frac{1}{2},\frac{1}{2},\frac{1}{4},\frac{3}{4} \\
        \end{array}
        \right)\ .
        \end{array}
        \end{equation}    
    
    2. For $X_{3,3}(1^6)$, we have $(a_1,a_2,a_3,a_4)=(1/3,1/3,2/3,2/3)$ and $\mu=3^6$.
    
    Family of solutions:
    \begin{equation}
        G_{4,4}^{m,n}\left((-1)^{m+n+4}\frac{w}{3^6}|
            \begin{array}{c}
             1,1,1,1 \\
             \frac{1}{3}',\frac{1}{3}',\frac{2}{3}',\frac{2}{3}' \\
            \end{array}
            \right)\ .
    \end{equation}
        
    Four linearly independent solutions:
    \begin{equation}\label{app:eqn_X33}
        \begin{array}{c}
            f_1(w)=G_{4,4}^{1,4}\left(-\frac{w}{3^6}|
            \begin{array}{c}
             1,1,1,1 \\
             \frac{1}{3},\frac{1}{3},\frac{2}{3},\frac{2}{3} \\
            \end{array}
            \right)\xrightarrow[]{|w|<3^6}\frac{ \Gamma \left(\frac{1}{3}\right)^4 }{ \Gamma \left(\frac{2}{3}\right)^2\Gamma\left(1\right)}\left(-\frac{w}{3^6}\right)^{1/3}\, _4F_3\left(\frac{1}{3},\frac{1}{3},\frac{1}{3},\frac{1}{3};\frac{2}{3},\frac{2}{3},1;\frac{w}{3^6}\right)\ ,\\
            \null\\
            f_2(w)=G_{4,4}^{1,4}\left(-\frac{w}{3^6}|
            \begin{array}{c}
             1,1,1,1 \\
             \frac{2}{3},\frac{1}{3},\frac{1}{3},\frac{2}{3} \\
            \end{array}
            \right)\xrightarrow[]{|w|<3^6}\frac{ \Gamma \left(\frac{2}{3}\right)^4 }{\Gamma\left(1\right)\Gamma \left(\frac{4}{3}\right)^2}\left(-\frac{w}{3^6}\right)^{2/3} \, _4F_3\left(\frac{2}{3},\frac{2}{3},\frac{2}{3},\frac{2}{3};1,\frac{4}{3},\frac{4}{3};\frac{w}{3^6}\right)\ ,\\
            \null\\
            f_3(w)=G_{4,4}^{2,4}\left(\frac{w}{3^6}|
            \begin{array}{c}
             1,1,1,1 \\
             \frac{1}{3},\frac{1}{3},\frac{2}{3},\frac{2}{3} \\
            \end{array}
            \right)\ ,\\
       \null\\
            f_4(w)=G_{4,4}^{2,4}\left(\frac{w}{3^6}|
            \begin{array}{c}
             1,1,1,1 \\
             \frac{2}{3},\frac{2}{3},\frac{1}{3},\frac{1}{3} \\
            \end{array}
            \right)\ .
        \end{array}
    \end{equation}
    
    3. For $X_{2,2,2,2}(1^8)$, we have $(a_1,a_2,a_3,a_4)=(1/2,1/2,1/2,1/2)$ and $\mu=2^8$.
    
    Family of solutions:
     \begin{equation}
         G_{4,4}^{m,n}\left((-1)^{m+n+4}\frac{w}{2^8}|
        \begin{array}{c}
         1,1,1,1 \\
         \frac{1}{2},\frac{1}{2},\frac{1}{2},\frac{1}{2} \\
        \end{array}
        \right)
        \ .\end{equation}
    
    Four linearly independent solutions:
        
        \begin{equation}\label{app:eqn_X2222}
            \begin{array}{c}
            f_1(w)=G_{4,4}^{1,4}\left(-\frac{w}{2^8}|
    \begin{array}{c}
     1,1,1,1 \\
     \frac{1}{2},\frac{1}{2},\frac{1}{2},\frac{1}{2} \\
    \end{array}
    \right)\xrightarrow[]{|w|<2^{8}}\frac{\Gamma\left(\frac{1}{2}\right)^4}{\Gamma\left(1\right)^3} \left(-\frac{w}{2^8}\right)^{1/2} \, _4F_3\left(\frac{1}{2},\frac{1}{2},\frac{1}{2},\frac{1}{2};1,1,1;\frac{w}{2^8}\right)\ ,\\
            \null\\
            f_2(w)=G_{4,4}^{2,4}\left(\frac{w}{2^8}|
    \begin{array}{c}
     1,1,1,1 \\
     \frac{1}{2},\frac{1}{2},\frac{1}{2},\frac{1}{2} \\
    \end{array}
    \right)\ ,\\
            \null\\
            f_3(w)=G_{4,4}^{3,4}\left(-\frac{w}{2^8}|
    \begin{array}{c}
     1,1,1,1 \\
     \frac{1}{2},\frac{1}{2},\frac{1}{2},\frac{1}{2} \\
    \end{array}
    \right)\ ,\\
            \null\\
            f_4(w)=G_{4,4}^{4,4}\left(\frac{w}{2^8}|
    \begin{array}{c}
     1,1,1,1 \\
     \frac{1}{2},\frac{1}{2},\frac{1}{2},\frac{1}{2} \\
    \end{array}
    \right)\ .
            \end{array}
        \end{equation}

\bibliographystyle{unsrtnat}
\bibliography{refs}

\begin{thebibliography}{41}
\providecommand{\natexlab}[1]{#1}
\providecommand{\url}[1]{\texttt{#1}}
\expandafter\ifx\csname urlstyle\endcsname\relax
  \providecommand{\doi}[1]{doi: #1}\else
  \providecommand{\doi}{doi: \begingroup \urlstyle{rm}\Url}\fi

\bibitem[Ooguri and Vafa(2007)]{Ooguri:2006in}
Hirosi Ooguri and Cumrun Vafa.
\newblock {On the Geometry of the String Landscape and the Swampland}.
\newblock \emph{Nucl. Phys.}, B766:\penalty0 21--33, 2007.
\newblock \doi{10.1016/j.nuclphysb.2006.10.033}.

\bibitem[Klaewer and Palti(2017)]{Klaewer:2016kiy}
Daniel Klaewer and Eran Palti.
\newblock {Super-Planckian Spatial Field Variations and Quantum Gravity}.
\newblock \emph{JHEP}, 01:\penalty0 088, 2017.
\newblock \doi{10.1007/JHEP01(2017)088}.

\bibitem[van Straten(2018)]{MR3822913}
Duco van Straten.
\newblock Calabi-{Y}au operators.
\newblock In \emph{Uniformization, {R}iemann-{H}ilbert correspondence,
  {C}alabi-{Y}au manifolds \& {P}icard-{F}uchs equations}, volume~42 of
  \emph{Adv. Lect. Math. (ALM)}, pages 401--451. Int. Press, Somerville, MA,
  2018.

\bibitem[Almkvist and von Straten()]{opcylist}
Gert Almkvist and Duco von Straten.
\newblock Calabi-yau differential operator database v.3.
\newblock URL \url{http://cydb.mathematik.uni-mainz.de/}.

\bibitem[Grimm et~al.(2018{\natexlab{a}})Grimm, Li, and Palti]{Grimm:2018cpv}
Thomas~W. Grimm, Chongchuo Li, and Eran Palti.
\newblock {Infinite Distance Networks in Field Space and Charge Orbits}.
\newblock 2018{\natexlab{a}}.

\bibitem[Klemm et~al.(1997)Klemm, Mayr, and Vafa]{Klemm:1996hh}
Albrecht Klemm, Peter Mayr, and Cumrun Vafa.
\newblock {BPS states of exceptional noncritical strings}.
\newblock \emph{Nucl. Phys. Proc. Suppl.}, 58:\penalty0 177, 1997.
\newblock \doi{10.1016/S0920-5632(97)00422-2}.
\newblock [,177(1996)].

\bibitem[Lee et~al.(2018)Lee, Lerche, and Weigand]{Lee:2018urn}
Seung-Joo Lee, Wolfgang Lerche, and Timo Weigand.
\newblock {Tensionless Strings and the Weak Gravity Conjecture}.
\newblock \emph{JHEP}, 10:\penalty0 164, 2018.
\newblock \doi{10.1007/JHEP10(2018)164}.

\bibitem[Klemm et~al.(1996)Klemm, Lerche, Mayr, Vafa, and Warner]{Klemm:1996bj}
Albrecht Klemm, Wolfgang Lerche, Peter Mayr, Cumrun Vafa, and Nicholas~P.
  Warner.
\newblock {Selfdual strings and N=2 supersymmetric field theory}.
\newblock \emph{Nucl. Phys.}, B477:\penalty0 746--766, 1996.
\newblock \doi{10.1016/0550-3213(96)00353-7}.

\bibitem[Seiberg and Witten(1994)]{Seiberg:1994rs}
N.~Seiberg and Edward Witten.
\newblock {Electric - magnetic duality, monopole condensation, and confinement
  in N=2 supersymmetric Yang-Mills theory}.
\newblock \emph{Nucl. Phys.}, B426:\penalty0 19--52, 1994.
\newblock \doi{10.1016/0550-3213(94)90124-4, 10.1016/0550-3213(94)00449-8}.
\newblock [Erratum: Nucl. Phys.B430,485(1994)].

\bibitem[Huang et~al.(2009{\natexlab{a}})Huang, Klemm, and
  Quackenbush]{Huang:2006hq}
Min-xin Huang, Albrecht Klemm, and Seth Quackenbush.
\newblock {Topological string theory on compact Calabi-Yau: Modularity and
  boundary conditions}.
\newblock \emph{Lect. Notes Phys.}, 757:\penalty0 45--102, 2009{\natexlab{a}}.
\newblock \doi{10.1007/978-3-540-68030-7_3}.

\bibitem[Blumenhagen et~al.(2018)Blumenhagen, Kl{\"a}wer, Schlechter, and
  Wolf]{Blumenhagen:2018nts}
Ralph Blumenhagen, Daniel Kl{\"a}wer, Lorenz Schlechter, and Florian Wolf.
\newblock {The Refined Swampland Distance Conjecture in Calabi-Yau Moduli
  Spaces}.
\newblock \emph{JHEP}, 06:\penalty0 052, 2018.
\newblock \doi{10.1007/JHEP06(2018)052}.

\bibitem[Grimm et~al.(2018{\natexlab{b}})Grimm, Palti, and
  Valenzuela]{Grimm:2018}
Thomas~W. Grimm, Eran Palti, and Irene Valenzuela.
\newblock {Infinite Distances in Field Space and Massless Towers of States}.
\newblock 2018{\natexlab{b}}.

\bibitem[Banks and Seiberg(2011)]{Banks:2010zn}
Tom Banks and Nathan Seiberg.
\newblock {Symmetries and Strings in Field Theory and Gravity}.
\newblock \emph{Phys. Rev.}, D83:\penalty0 084019, 2011.
\newblock \doi{10.1103/PhysRevD.83.084019}.

\bibitem[Candelas and de~la Ossa(1991)]{CANDELAS1991455}
Philip Candelas and Xenia~C. de~la Ossa.
\newblock Moduli space of calabi-yau manifolds.
\newblock \emph{Nuclear Physics B}, 355\penalty0 (2):\penalty0 455 -- 481,
  1991.
\newblock ISSN 0550-3213.
\newblock \doi{https://doi.org/10.1016/0550-3213(91)90122-E}.
\newblock URL
  \url{http://www.sciencedirect.com/science/article/pii/055032139190122E}.

\bibitem[Doran and Morgan(2006)]{MR2282973}
Charles~F. Doran and John~W. Morgan.
\newblock Mirror symmetry and integral variations of {H}odge structure
  underlying one-parameter families of {C}alabi-{Y}au threefolds.
\newblock In \emph{Mirror symmetry. {V}}, volume~38 of \emph{AMS/IP Stud. Adv.
  Math.}, pages 517--537. Amer. Math. Soc., Providence, RI, 2006.

\bibitem[Morrison(1992)]{MR1191426}
David~R. Morrison.
\newblock Picard-{F}uchs equations and mirror maps for hypersurfaces.
\newblock In \emph{Essays on mirror manifolds}, pages 241--264. Int. Press,
  Hong Kong, 1992.

\bibitem[Klemm and Theisen(1993)]{Klemm:1992tx}
Albrecht Klemm and Stefan Theisen.
\newblock {Considerations of one modulus Calabi-Yau compactifications:
  Picard-Fuchs equations, Kahler potentials and mirror maps}.
\newblock \emph{Nucl. Phys.}, B389:\penalty0 153--180, 1993.
\newblock \doi{10.1016/0550-3213(93)90289-2}.

\bibitem[Font(1993)]{Font:1992uk}
Anamaria Font.
\newblock {Periods and duality symmetries in Calabi-Yau compactifications}.
\newblock \emph{Nucl. Phys.}, B391:\penalty0 358--388, 1993.
\newblock \doi{10.1016/0550-3213(93)90152-F}.

\bibitem[Libgober and Teitelbaum(1993)]{MR1201748}
A.~Libgober and J.~Teitelbaum.
\newblock Lines on {C}alabi-{Y}au complete intersections, mirror symmetry, and
  {P}icard-{F}uchs equations.
\newblock \emph{Internat. Math. Res. Notices}, \penalty0 (1):\penalty0 29--39,
  1993.
\newblock ISSN 1073-7928.
\newblock \doi{10.1155/S1073792893000030}.
\newblock URL \url{https://doi.org/10.1155/S1073792893000030}.

\bibitem[Klemm and Theisen(1994)]{Klemm:1993jj}
Albrecht Klemm and Stefan Theisen.
\newblock {Mirror maps and instanton sums for complete intersections in
  weighted projective space}.
\newblock \emph{Mod. Phys. Lett.}, A9:\penalty0 1807--1818, 1994.
\newblock \doi{10.1142/S0217732394001660}.

\bibitem[Cox and Katz(1999)]{MR1677117}
David~A. Cox and Sheldon Katz.
\newblock \emph{Mirror symmetry and algebraic geometry}, volume~68 of
  \emph{Mathematical Surveys and Monographs}.
\newblock American Mathematical Society, Providence, RI, 1999.
\newblock ISBN 0-8218-1059-6.
\newblock \doi{10.1090/surv/068}.
\newblock URL \url{https://doi.org/10.1090/surv/068}.

\bibitem[Hosono et~al.(1995)Hosono, Klemm, Theisen, and Yau]{Hosono:1994ax}
S.~Hosono, A.~Klemm, S.~Theisen, and Shing-Tung Yau.
\newblock {Mirror symmetry, mirror map and applications to complete
  intersection Calabi-Yau spaces}.
\newblock \emph{Nucl. Phys.}, B433:\penalty0 501--554, 1995.
\newblock \doi{10.1016/0550-3213(94)00440-P}.
\newblock [AMS/IP Stud. Adv. Math.1,545(1996)].

\bibitem[Schmid(1973)]{MR0382272}
Wilfried Schmid.
\newblock Variation of {H}odge structure: the singularities of the period
  mapping.
\newblock \emph{Invent. Math.}, 22:\penalty0 211--319, 1973.
\newblock ISSN 0020-9910.
\newblock \doi{10.1007/BF01389674}.
\newblock URL \url{https://doi.org/10.1007/BF01389674}.

\bibitem[Wang(1997)]{wang1997incompleteness}
Chin-Lung Wang.
\newblock On the incompleteness of the weil-petersson metric along
  degenerations of calabi-yau manifolds.
\newblock \emph{Mathematical Research Letters}, 4\penalty0 (1):\penalty0
  157--171, 1997.

\bibitem[Greene and Plesser(1990)]{GREENE199015}
B.R. Greene and M.R. Plesser.
\newblock Duality in calabi-yau moduli space.
\newblock \emph{Nuclear Physics B}, 338\penalty0 (1):\penalty0 15 -- 37, 1990.
\newblock ISSN 0550-3213.
\newblock \doi{https://doi.org/10.1016/0550-3213(90)90622-K}.
\newblock URL
  \url{http://www.sciencedirect.com/science/article/pii/055032139090622K}.

\bibitem[{Libgober} and {Teitelbaum}(1993)]{1993alg.geom..1001L}
A.~{Libgober} and J.~{Teitelbaum}.
\newblock {Lines on Calabi Yau complete intersections, mirror symmetry, and
  Picard Fuchs equations}.
\newblock In \emph{eprint arXiv:alg-geom/9301001}, January 1993.

\bibitem[Berglund et~al.(1994)Berglund, Candelas, de~la Ossa, Font, H{\"u}bsch,
  Jan{\v c}i{\'c}, and Quevedo]{Berglund94}
Per Berglund, Philip Candelas, Xenia de~la Ossa, Anamar{\'\i}a Font, Tristan
  H{\"u}bsch, Dubravka Jan{\v c}i{\'c}, and Fernando Quevedo.
\newblock Periods for calabi-yau and landau-ginzburg vacua.
\newblock \emph{Nuclear Physics B}, 419\penalty0 (2):\penalty0 352 -- 403,
  1994.
\newblock ISSN 0550-3213.
\newblock \doi{https://doi.org/10.1016/0550-3213(94)90047-7}.
\newblock URL
  \url{http://www.sciencedirect.com/science/article/pii/0550321394900477}.

\bibitem[Klemm et~al.()Klemm, Scheidegger, and Zagier]{Klemm18}
Albrecht Klemm, Emanuel Scheidegger, and Don Zagier.
\newblock Periods and quasiperiods of modular forms and d-brane masses for the
  mirror quintic.
\newblock \emph{in preparation}.

\bibitem[Antoniadis et~al.(1994)Antoniadis, Gava, Narain, and
  Taylor]{Antoniadis:1993ze}
Ignatios Antoniadis, E.~Gava, K.~S. Narain, and T.~R. Taylor.
\newblock {Topological amplitudes in string theory}.
\newblock \emph{Nucl. Phys.}, B413:\penalty0 162--184, 1994.
\newblock \doi{10.1016/0550-3213(94)90617-3}.

\bibitem[Gopakumar and Vafa(1998)]{Gopakumar:1998jq}
Rajesh Gopakumar and Cumrun Vafa.
\newblock {M theory and topological strings. 2.}
\newblock 1998.

\bibitem[Katz et~al.(1999)Katz, Klemm, and Vafa]{Katz:1999xq}
Sheldon~H. Katz, Albrecht Klemm, and Cumrun Vafa.
\newblock {M theory, topological strings and spinning black holes}.
\newblock \emph{Adv. Theor. Math. Phys.}, 3:\penalty0 1445--1537, 1999.
\newblock \doi{10.4310/ATMP.1999.v3.n5.a6}.

\bibitem[Pandharipande and Thomas(2010)]{MR2552254}
R.~Pandharipande and R.~P. Thomas.
\newblock Stable pairs and {BPS} invariants.
\newblock \emph{J. Amer. Math. Soc.}, 23\penalty0 (1):\penalty0 267--297, 2010.
\newblock ISSN 0894-0347.
\newblock \doi{10.1090/S0894-0347-09-00646-8}.
\newblock URL \url{https://doi.org/10.1090/S0894-0347-09-00646-8}.

\bibitem[Strominger(1995)]{Strominger:1995cz}
Andrew Strominger.
\newblock {Massless black holes and conifolds in string theory}.
\newblock \emph{Nucl. Phys.}, B451:\penalty0 96--108, 1995.
\newblock \doi{10.1016/0550-3213(95)00287-3}.

\bibitem[Huang et~al.(2009{\natexlab{b}})Huang, Klemm, Marino, and
  Tavanfar]{Huang:2007sb}
Min-xin Huang, Albrecht Klemm, Marcos Marino, and Alireza Tavanfar.
\newblock {Black holes and large order quantum geometry}.
\newblock \emph{Phys. Rev.}, D79:\penalty0 066001, 2009{\natexlab{b}}.
\newblock \doi{10.1103/PhysRevD.79.066001}.

\bibitem[Vafa(1995)]{Vafa:1995ta}
Cumrun Vafa.
\newblock {A Stringy test of the fate of the conifold}.
\newblock \emph{Nucl. Phys.}, B447:\penalty0 252--260, 1995.
\newblock \doi{10.1016/0550-3213(95)00279-2}.

\bibitem[Greene(1996)]{Greene:1996cy}
Brian~R. Greene.
\newblock {String theory on Calabi-Yau manifolds}.
\newblock In \emph{{Fields, strings and duality. Proceedings, Summer School,
  Theoretical Advanced Study Institute in Elementary Particle Physics, TASI'96,
  Boulder, USA, June 2-28, 1996}}, page~64, 1996.

\bibitem[Candelas et~al.(1991)Candelas, De~La~Ossa, Green, and
  Parkes]{Candelas:1990rm}
Philip Candelas, Xenia~C. De~La~Ossa, Paul~S. Green, and Linda Parkes.
\newblock {A Pair of Calabi-Yau manifolds as an exactly soluble superconformal
  theory}.
\newblock \emph{Nucl. Phys.}, B359:\penalty0 21--74, 1991.
\newblock \doi{10.1016/0550-3213(91)90292-6}.
\newblock [AMS/IP Stud. Adv. Math.9,31(1998)].

\bibitem[Ceresole et~al.(1995)Ceresole, D'Auria, Ferrara, and
  Van~Proeyen]{Ceresole:1995jg}
Anna Ceresole, R.~D'Auria, S.~Ferrara, and Antoine Van~Proeyen.
\newblock {Duality transformations in supersymmetric Yang-Mills theories
  coupled to supergravity}.
\newblock \emph{Nucl. Phys.}, B444:\penalty0 92--124, 1995.
\newblock \doi{10.1016/0550-3213(95)00175-R}.

\bibitem[{Meijer}(1936)]{zbMATH02526339}
C.~S. {Meijer}.
\newblock {\"Uber Whittakersche bzw. Besselsche Funktionen und deren Produkte.}
\newblock \emph{{Nieuw Arch. Wiskd., II. Ser.}}, 18\penalty0 (4):\penalty0
  10--39, 1936.
\newblock ISSN 0028-9825.

\bibitem[Beals and Szmigielski(2013)]{beals2013meijer}
Richard Beals and Jacek Szmigielski.
\newblock Meijer g-functions: a gentle introduction.
\newblock \emph{Notices of the AMS}, 60\penalty0 (7):\penalty0 866--872, 2013.

\bibitem[Bateman(1953)]{bateman1953higher}
Harry Bateman.
\newblock Higher transcendental functions [volume i], p. 206, 1953.

\end{thebibliography}








	\end{document}